\documentclass[twocolumn,showpacs,showkeys,amsmath,amssymb,prd,10pt]{revtex4}

\usepackage{graphicx}% Include figure files
\usepackage{dcolumn}% Align table columns on decimal point
\usepackage{bm}% bold math
\usepackage{hyperref}
\usepackage{color}

\definecolor{Blue}{rgb}{0.3,0.3,0.9}
\definecolor{Red}{rgb}{1,0,0}
\definecolor{Green}{rgb}{0,1,0}

\def\smmath#1{\scriptstyle #1}

\begin{document}
\title{Observing the Big Bounce with Tensor Modes in the Cosmic Microwave Background:
Phenomenology and Fundamental LQC Parameters}

\author{Julien Grain}
 \email{julien.grain@ias.u-psud.fr}
 \affiliation{%
 Univ. Paris-Sud, Institut d'Astrophysique Spatiale, UMR8617, Orsay, France, F-91405}
\affiliation{%
CNRS, Orsay, France, F-91405}
\author{Aur\'elien Barrau}%
 \email{aurelien.barrau@cern.ch}
 \author{ Thomas Cailleteau}%
 \email{cailleteau@lpsc.in2p3.fr}
\affiliation{%
Laboratoire de Physique Subatomique et de Cosmologie, UJF, INPG, CNRS, IN2P3\\
53, avenue des Martyrs, 38026 Grenoble cedex, France
}%
\author{Jakub Mielczarek}%
 \email{jakub.mielczarek@uj.edu.pl}
\affiliation{%
Astronomical Observatory, Jagiellonian University, 30-244
Krak\'ow, Orla 171, Poland}

\date{\today}

\begin{abstract}

Cosmological models where the standard big bang is replaced by 
a bounce have been studied for decades. The situation has, however,
dramatically changed in the past years for two reasons: first, because
new  ways to probe the early Universe have emerged, in
particular, thanks to the cosmic microwave background, and second, 
because some well grounded theories -- especially loop quantum 
cosmology -- unambiguously predict a bounce, at least for homogeneous models. 
In this article, we investigate 
into the details the phenomenological parameters that could be constrained 
or measured by next-generation B-mode cosmic micorwave background experiments. We point 
out that an important observational window could be opened. We then 
show that those constraints can be converted into very meaningful limits 
on the fundamental loop quantum cosmology parameters. This establishes
the early Universe as an invaluable quantum gravity laboratory.

\end{abstract}

%Use showkeys class option if keyword

\maketitle

\section{Introduction}
%\paragraph*{Introduction.}
The big bang paradigm is unquestionably a major achievement 
of contemporary science. However, in parallel to its successes it 
raises some very fundamental questions. Among them are of course 
the dark matter and dark energy issues. Nevertheless, the big bang 
singularity  remains, in itself, one of the greatest puzzles of the 
whole approach. It is a nearly unavoidable prediction of general 
relativity where the theory is, precisely, not correct anymore. Solving
the singularity by replacing the big bang by a big bounce is one of
the main achievement of loop quantum cosmology (LQC) 
\cite{lqc_review} as a symmetry reduced version  of the loop quantum
gravity (LQG) scheme to nonperturbatively quantize general 
relativity in a background invariant way \cite{lqg_review}. 

Moreover, if the Universe is assumed to be filled with a scalar field
described by a self-interaction potential well, the contracting phase -- preceding the Big
Bounce -- can set the field in the appropriate conditions for a phase of slow-roll 
inflation to start just after the bounce. In the specific (and simple) case of a massive scalar 
field, and in the framework of an effective LQC universe, only a tiny amount of 
potential energy at the bounce is necessary for a long enough phase of inflation 
to be {\it naturally} generated \cite{jakub2010a,jakub2010b,ashtekar2010}. 
In effective LQC, it is therefore possible both to solve 
the big bang singularity and to generate the specific conditions necessary for inflation to take
place. 
Finally, and this is the keypoint addressed by this paper,
such a model can, in principle, be tested. The quantum fluctuations leading to the
cosmological perturbations observed in the cosmic microwave background (CMB) anisotropies, 
though still stretched to astronomical size by inflation, experienced the influence of the 
contraction phase and of the bounce. As a consequence, the statistical properties of 
cosmological perturbations are potentially distorted as compared to the standard inflationary 
prediction. This finally translates into distortions in the angular power spectra of CMB 
anisotropies.

Up to now, only  corrections to tensor modes of the cosmological 
perturbations have been rigorously derived in LQC \cite{Bojowald:2007cd}, 
potentially leaving a footprint on the CMB B-mode. Although not 
yet detected and marginally within the aims of the Planck satellite, 
the measurement of B-mode polarization will be the core of the future 
CMBPol/B-Pol missions \cite{cmbpol}. We therefore investigate a 
possible detection of the big bounce using future B-mode 
measurements, by considering first the phenomenological aspects
and then turning to the fundamental parameters.

Our paper is organized as follows. In Sec. \ref{sec:bmode}, we describe how the B-mode power 
spectrum is distorted in a \{bouncing+inflationary\} universe as compared to the standard 
prediction from inflation and argue that this distortion can be parametrized by two 
phenomenological parameters (denoted $k_\star$ and $R$) encoding the properties of the bounce. 
The question of a possible detection of the bounce with the B-mode angular
power spectrum translates into the determination of specific values of the two 
phenomenological parameters describing the distorted shape of the tensor power spectrum. 
Then, Sec. \ref{sec:fisher} is devoted to a brief presentation of the Fisher analysis we have
used 
to define the signal-to-noise ratio associated with the cosmological parameters shaping the 
B-mode power spectrum. We apply this approach to the specific case of $k_\star$ and $R$, 
assuming the experimental characteristics of the future CMBPol/B-Pol missions and present our 
numerical results in Sec. \ref{sec:pheno}. In Sec. \ref{sec:fund}, the range of 
phenomenological parameters leading to a possible detection is translated into possible values 
of the more fundamental LQC parameters. We finally discuss our results and conclude 
 in Sec. \ref{sec:concl}.

\section{CMB B-mode with a bounce}
\label{sec:bmode}
\subsection{Primordial power spectrum for tensor modes}
\label{sec:Pt}
Many articles \cite{lqcgen} 
have been devoted to the study of gravitational waves in LQC. 
We focus in this paper on the simplest (and, in our opinion, most 
convincing) scenario (essentially developed in \cite{jakub2010a,jakub2010b}): a universe filled with a single massive scalar 
field. This accounts impressively well for the observed Universe. 
Before the bounce, the Hubble parameter is negative (therefore 
acting as an antifriction term) and makes the field climb up its 
potential. After the bounce, the Hubble constant becomes positive (therefore 
a friction term) and naturally leads to a standard phase of slow-roll 
inflation. It is remarkable that inflation naturally occurs without
any fine tuning.

The main characteristics of a "bouncy" power spectrum for tensor modes are the following:
\begin{itemize}
\item The IR part is $k^2$ suppressed. This is due to the freezing of very large-scale modes in 
the Minkowski vacuum. Those modes indeed exit the horizon long {\it before} the bounce and
naturally exhibit a quadratic spectrum.
\item The UV part is identical to the standard prediction. Small scales indeed 
experience a history basically similar to that of the big bang scenario. They exit the horizon
during inflation and reenter later, leading to the standard nearly scale-invariant spectrum.
\item Intermediate scales, around $k\approx k_\star$, exhibit both a bump of 
amplitude $R$ and damped oscillations. This is mostly due to the fact that all modes are 
inevitably in causal contact at the bounce (the Hubble parameter vanishes, therefore leading
to an infinite Hubble radius). 
\end{itemize}
Those characteristics have been fully determined by numerically solving the equations of motion 
of tensor perturbations with LQC corrections  propagating in a \{bouncing+inflationary\} 
universe \cite{jakub2010b}. It is worth underlining that those equations of motion, as obtained in 
Ref. \cite{Bojowald:2007cd}, are derived from an algebra which is anomaly-free at all orders and 
can be safely used throughout the 
entire history of the bouncing universe. This may not be true anymore with scalar
perturbations.

In our previous work \cite{jakub2010b}, two possible phenomenological descriptions of the  
primordial tensor power spectrum have been proposed. The first, and more complicated, description 
introduces three phenomenological parameters to approximate the shape of the time-dependent 
effective mass of gravity waves propagating in the LQC universe. It captures all the detailed
characteristics of the primordial power spectrum. The interested reader is referred to Sec. IV 
of Ref. \cite{jakub2010b} for a detailed discussion.

The second, and simpler one, is summarized by the following equation:
\begin{equation}
\mathcal{P}_{\text{T}}= \frac{16}{\pi} \left(\frac{H}{m_{\text{Pl}}} \right)^2 
\frac{\left( \frac{k}{aH}\right)^{n_\mathrm{T}} }{1+(k_\star/k)^2} 
\left[1+\frac{4R-2}{1+(k/k_\star)^2} \right],  \label{eff1}
\end{equation}
where $H$ is the Hubble constant at horizon crossing {\it after} the bounce. It is more than
enough to compute potentially observable effects. In the above formula,
$$
\mathcal{P}^{STD}_{\mathrm{T}}\equiv\frac{16}{\pi} \left(\frac{H}{m_{\text{Pl}}} \right)^2\left( \frac{k}{aH}\right)^{n_\mathrm{T}} 
$$
stands for the power spectrum corresponding to the standard inflationary universe while
$$
\frac{\mathcal{P}_{\text{T}}}{\mathcal{P}^{STD}_{\mathrm{T}}}=\frac{1}{1+(k_\star/k)^2} 
\left[1+\frac{4R-2}{1+(k/k_\star)^2} \right]
$$
corresponds to the LQC corrections. This spectrum is completely determined by four parameters: $R$ and $k_\star$, 
encoding the LQC corrections, the spectral index $n_\mathrm{T}$, and the normalization, given by the tensor-to-scalar ratio $T/S$ defined in the 
UV limit. In the following, the values chosen for $T/S$ correspond to an amplitude of the scalar 
perturbations given by the WMAP 7-yr best fit, {\it i.e.} 
$\mathcal{A}_\text{S}\simeq2.49\times10^{-9}$. 
Though this value assumes a power-law shape for the scalar power spectrum (which is not 
guaranteed in a bouncy universe), this is only a matter of convention and any change in 
$\mathcal{A}_\text{S}$ can  be reabsorbed in a new convention for $T/S$. Nevertheless, 
this choice makes sense in the UV limit and allows us to remain consistent with the standard
B-mode parametrization. The damped oscillations are approximated by an envelope function and 
$k_\star$ is simply interpreted as the wavenumber 
associated with the modes crossing out the horizon when the phase of slow-roll inflation starts. 
This parameter will therefore decrease as the number of e-folds of inflation increases.

\subsection{B-mode angular power spectrum}

The  B-mode angular power spectrum is made of two components:
\begin{itemize}
\item the primordial part, due to gravity waves produced in the early Universe, denoted $C^{B,prim}_\ell$ in the following and
\item the secondary component, due to lensing converting E-mode into B-mode, denoted $C^{B,lens}_\ell$.
\end{itemize}

\subsubsection{Primordial component}
The shape of the primordial part of $C^B_\ell$ is driven both by the phenomenological 
parameters describing the primordial tensor power spectrum 
($k_\star,~R,~n_\mathrm{T}$ and $T/S$ if one uses Eq. (\ref{eff1}) to parametrize 
$\mathcal{P}_\mathrm{T}$) and by standard cosmological parameters 
(in particular $\Omega_\Lambda,~\Omega_{CDM},~\Omega_k$ and the optical depth to reionization $\tau$). 

\begin{figure}
\begin{center}
\includegraphics[scale=0.48]{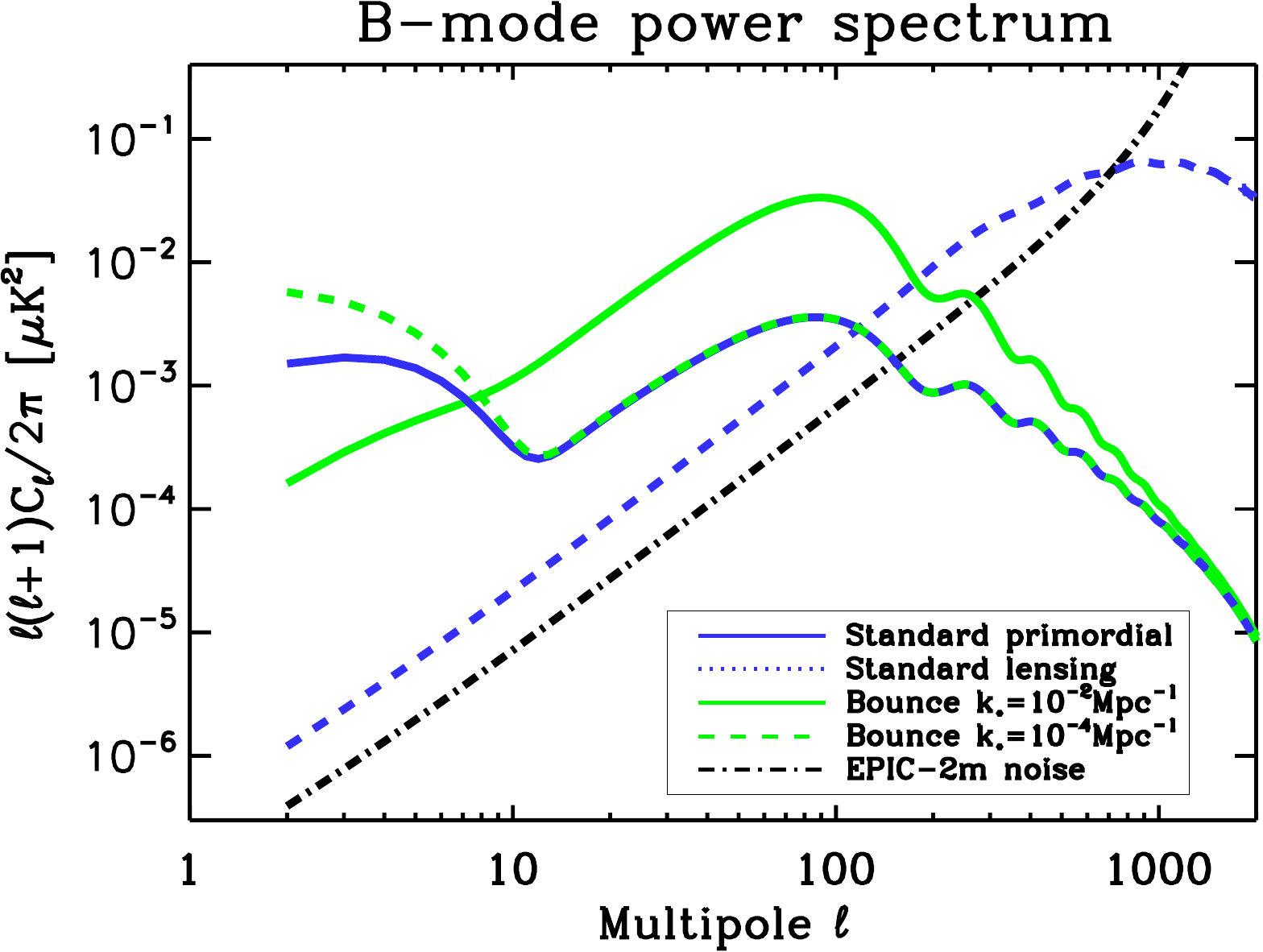}
\caption{Standard (blue curve) and typical bounce-modified (green curve) 
$C^B_\ell$ spectra for two values of $k_\star$. Other cosmological 
parameters are given by WMAP 7-yrs best fit plus $R=10,~n_\mathrm{T}=-0.012,~T/S=0.05$ and $\tau=0.087$.}
\label{cl}
\end{center}
\end{figure}

First of all, to understand qualitatively how LQC modifies the B-mode angular power spectrum, $C^{B,LQC}_\ell$
can be roughly approximated by
\begin{eqnarray}
	C^{B,LQC}_{\ell}=\frac{C^{B,STD}_\ell}{1+\left(\ell_\star/\ell\right)}\left[1+\frac{4R-2}{1+\left(\ell/\ell_\star\right)}\right].
	\label{eff2}
\end{eqnarray}
In the above, $\ell_\star=k_\star/k_H$, where $k_H\sim2.3\times10^{-4}$ Mpc$^{-1}$ is
the Hubble wavenumber today, and $C^{B,STD}_\ell$ stands for the B-mode power spectrum as 
obtained without LQC corrections ({\it i.e.}, the B-mode obtained by assuming the standard 
power law for the primordial tensor power spectrum parametrized with $n_\mathrm{T}$ and $T/S$). 
From this simple parametrization, two regimes can easily be identified, depending on the value 
of $k_\star/k_H$. For $k_\star/k_H>1$, the LQC B-mode power spectrum exhibits
\begin{itemize}
\item a suppression of power for $\ell<\ell_\star$ and 
\item a bump around $\ell\sim\ell_\star$ and
\item coi{n}cides with the standard inflationary prediction for $\ell>\ell_\star$. 
\end{itemize}
For $k_\star/k_H<1$, the IR suppression corresponds to length scales which are much greater 
than the observable scales and the LQC-corrected B-mode power spectrum
\begin{itemize}
\item exhibits a boost of  power at large angular scales corresponding to the tail of the bump in $\mathcal{P}_\mathrm{T}$ and
\item coincides with the standard inflationary prediction at intermediate and small angular scales.
\end{itemize}

To perform a more detailed analysis, the phenomenological spectra presented in Sec. \ref{sec:Pt}
have then been used as inputs for the 
primordial tensor perturbations and then converted into observable $C^B_\ell$ 
spectra by using {\sc Camb} \cite{camb}. Figure~\ref{cl} gives an example of how the angular 
power spectrum is distorted due to the bounce using Eq. (\ref{eff1}) as an input for the 
tensor spectrum and assuming two different values of the "transition" length 
scale $k_\star=10^{-4}$ and $10^{-2}$~Mpc${}^{-1}$. These numerically computed B-mode power 
spectra are not fundamentally different from the zeroth order approximation of $C^B_\ell$ given 
in Eq. (\ref{eff2}), although they show some slightly different features.

\begin{figure}
\begin{center}
\includegraphics[scale=0.48]{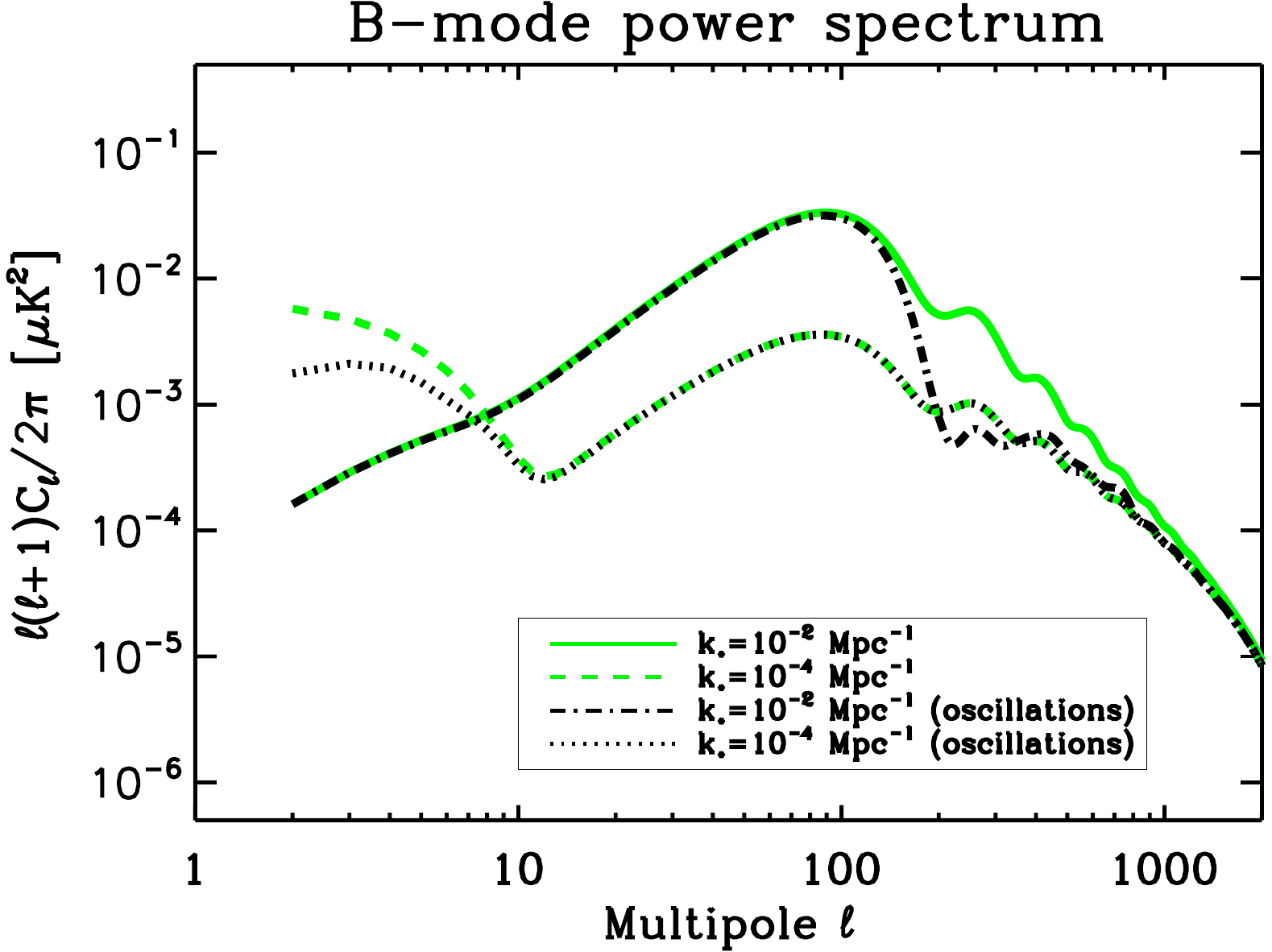}
\caption{B-mode power spectrum computed with (black curves) and without (green curves) oscillations in the bouncy primordial power spectrum of tensor modes for two values of $k_\star$. (Other cosmological parameters are as in Fig. \ref{cl}.)}
\label{closcil}
\end{center}
\end{figure}

In Fig.~\ref{closcil}, the resulting B-mode spectra with and without the damped oscillations 
are displayed for the same values of the transition length scale $k_\star$. For $k_\star<k_H$, 
neglecting the damped oscillations in $\mathcal{P}_{\mathrm{T}}$ leads to an 
overestimation of the boost at large angular scales. For $k_\star>k_H$, using Eq. (\ref{eff1}) 
results in an overestimation of the power just after the bump located at $\ell_\star$. The
effects of oscillations are always small (the IR suppression and the bump at $k_\star$ are by
far the more important observational features) and can be accounted for in Eq. (\ref{eff1}) by 
just considering an effective bump  $R_{\text{eff}}$ slightly smaller than $R$ for $k>k_\star$.

\begin{figure*}
\begin{center}
\includegraphics[scale=0.48]{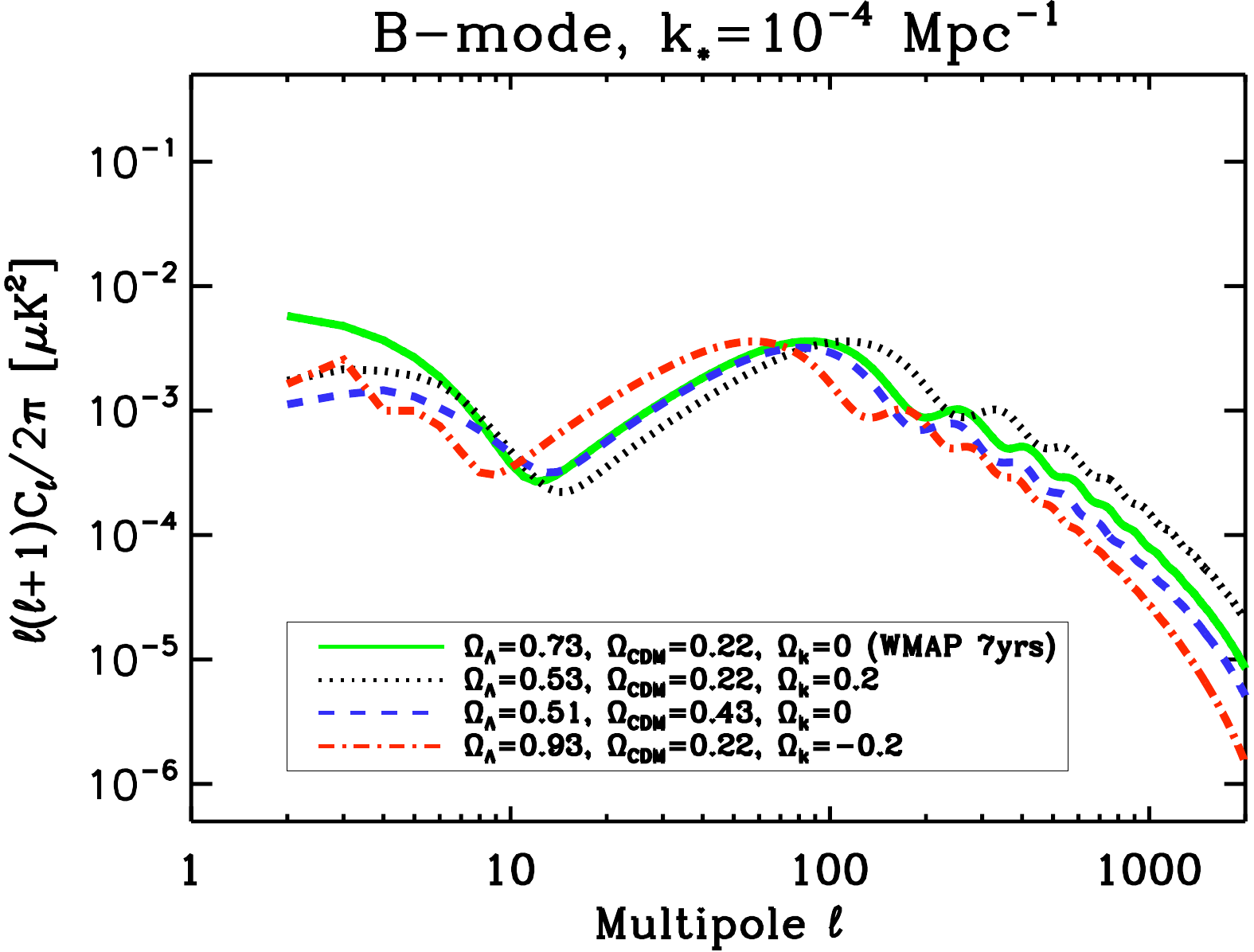} \hspace{0.5cm} \includegraphics[scale=0.48]{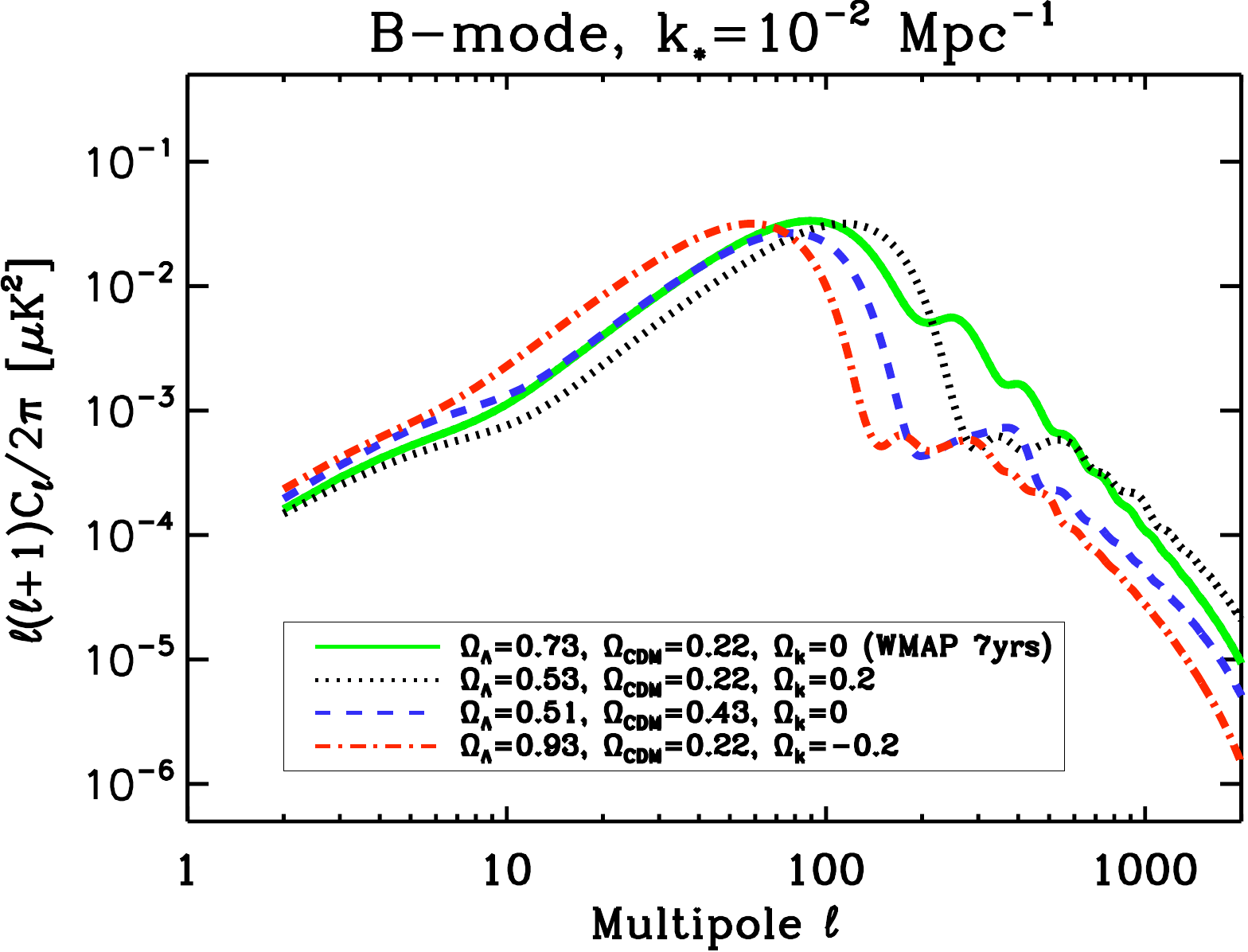}
\caption{B-mode power spectrum computed for different values of $\Omega_\Lambda,~\Omega_{CDM}$ and $\Omega_k$ and for $k_\star=10^{-4}$~Mpc${}^{-1}$ (left panel) and $k_\star=10^{-2}$~Mpc${}^{-1}$ (right panel). Other cosmological parameters are $k_\star=10^{-2}$~Mpc${}^{-1}$, $R=10,~n_\mathrm{T}=-0.012,~T/S=0.05$ and $\tau=0.087$.}
\label{clcosmo}
\end{center}
\end{figure*}

In Fig.~(\ref{clcosmo}), the primordial B-mode power spectrum is shown for different values of 
$\Omega_\Lambda,~\Omega_{CDM}$, and $\Omega_k$ and for two values of $k_\star$. For $k_\star>k_H$, 
the main effect is a shift in the overall power spectrum  without changing its shape. 
For $k_\star<k_H$, varying the parameters leads to a shift 
in $\ell$ for multipoles greater than $\sim10$ and to a slight suppression of power for 
$\ell<10$. 

Finally the primordial part of the B-mode angular power spectrum is also shaped by the optical depth 
to reionization $\tau$. The associated impact can be inferred from the simple expression
given by Eq. (\ref{eff2}) and is similar to what happens in the standard 
case. Reionization leads first to a boost of power a large angular scales, roughly scaling as 
$(1-e^{-\tau})^2$, and second, to a slight suppression at smaller angular scales scaling as 
$e^{-2\tau}$.

\subsubsection{Lensing component}
The lensing part of the CMB B-mode being given by the convolution of the E-mode power spectrum 
with the deflection field power spectrum, its computation implicitly assumes that the 
primordial power spectrum of {\it scalar} perturbations is known. Unfortunately, the LQC-corrected scalar power spectrum is still being debated and the exact shape of the secondary component 
of the B-mode cannot be {\it a priori} safely computed. However, this secondary component will 
be considered as a nuisance parameter ({\it i.e.} as an additional noise) spoiling the 
primordial component used to estimate the cosmological parameters. As a 
consequence, a reasonable estimate of the lensing B-mode is sufficient to investigate
the detectability of LQC parameters by using the CMB signal generated by primordial 
gravity waves.

The lensing B-mode without LQC correction is fixed by our theoretical knowledge of the 
deflection field and by our {\it observational} knowledge of the temperature (denoted 
T-mode hereafter) and E-mode angular power spectra of the CMB. Any strong 
modifications of the lensing B-mode power spectrum therefore implicitly assume strong 
distortions of the  T-mode and E-mode angular power spectra. As those spectra are well measured
it is not worth considering a substantial modification of the lensing component. 
This would anyway be a
subdominant effect when compared to other uncertainties.

%From this simple statement, we can infer that the 
%primordial scalar power spectrum should not be modified too much by LQC, otherwise the model 
%would be already excluded. (Fancy combinations of 'exotic' values of the cosmological 
%parameters with a strong modification of the scalar power spectrum may turn out to be 
%consistent with current T- and E-mode measurements. However, such an option seems highly 
%unlikely.) More specifically, it appears reasonable to consider that such primordial power 
%spectrum for scalar perturbations is not modified, as compared to the standard prediction, 
%for $k>k_H$. Lensing B-mode being a nuisance signal, the worth case would be a strong boost of 
%the primordial scalar power spectrum for $k<k_H$, translating into a boost of $C^{B,lens}_\ell$ 
%at large angular scales. (We stress out that such an hypothetical boost is however unfavored by 
%the last WMAP measurements of $C^T_\ell$ at large angular scales).

\begin{figure}
\begin{center}
\includegraphics[scale=0.48]{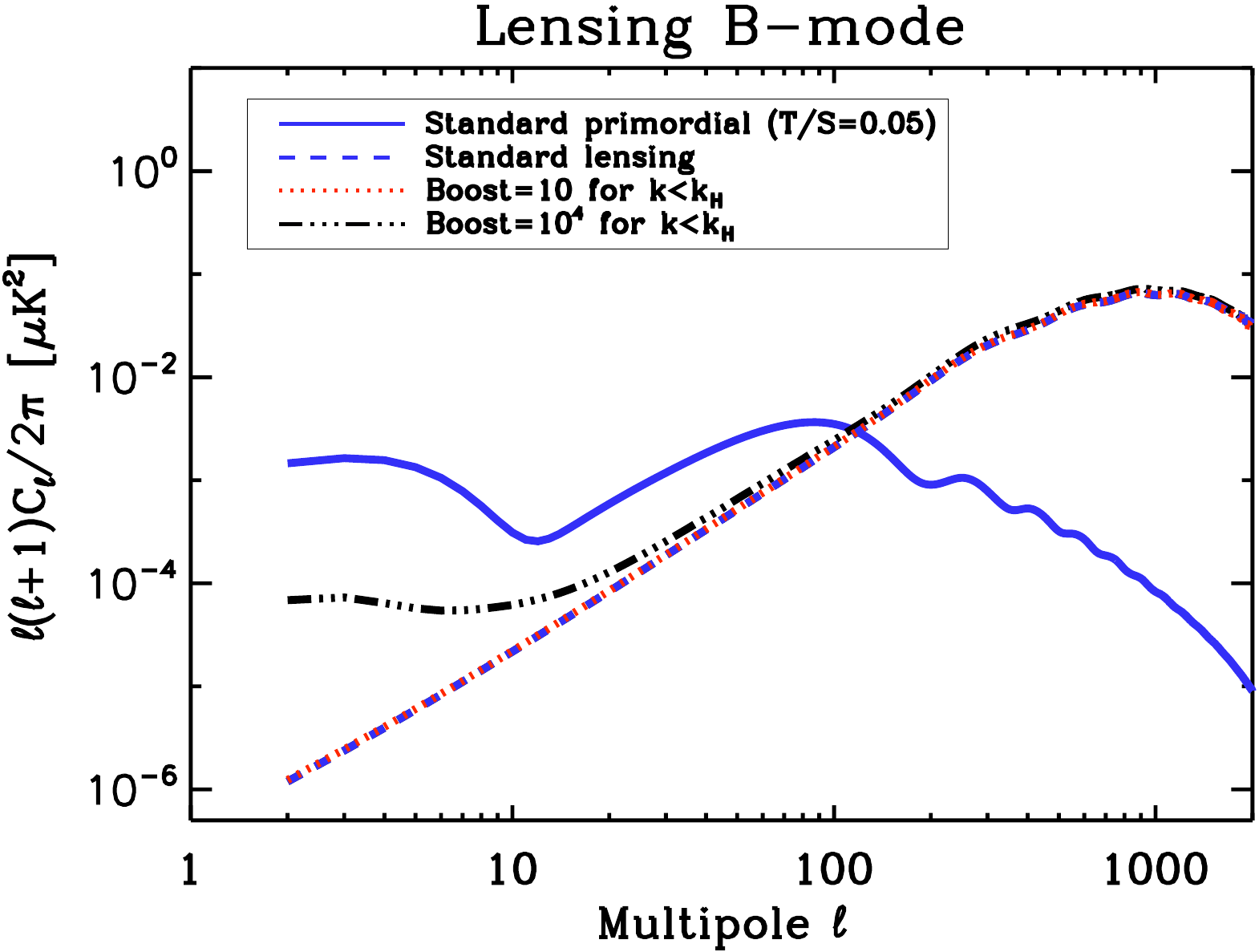}
\caption{Lensing-induced B-mode assuming that scalar perturbations might be boosted for $k<k_H$.}
\label{cllensing}
\end{center}
\end{figure}

Some $C^{B,lens}_\ell$ spectra are displayed in Fig. \ref{cllensing}. One is simply derived from
the standard 
inflationary prediction, the amplitude and spectral index of the scalar perturbations being 
fixed to their WMAP 7-yr best fit values, and the others are obtained by boosting the 
primordial scalar power 
spectrum for wavenumbers smaller than the Hubble scale. It clearly shows that as long as 
unrealistic values of the boost ({\it e.g.}, 10,000) are not considered, the shape of the 
lensing-induced B-mode power spectrum can safely be fixed to its standard prediction.

\section{Detecting the bounce: the strategy}
\label{sec:fisher}
\subsection{Parametrizing the B-mode power spectrum}
In view of the previous results, the primordial component of the B-mode angular power spectrum 
is determined by the five following parameters: $k_\star,~R,~n_\mathrm{T},~T/S $ and $\tau$, 
denoted $\theta_i$ hereafter. The other cosmological parameters will be fixed to the WMAP 7-yr 
best fit, and the lensing-induced B-mode will be fixed to its standard prediction. We will also 
neglect the effect of damped oscillations. The effect of  oscillations on $C^{B,prim}_\ell$ can anyway be recasted in the "language" of 
Eq. (\ref{eff1}) by introducing an effective bump $R_{\text{eff}}$. Using Eq. (\ref{eff1}) to
parametrize the primordial tensor power spectrum therefore provides a reliable  
description of the physics at play in LQC in all cases by considering that the detectable values of $R$ are
to be interpreted as an effective bump.

%As explained before, fixing all the cosmological parameters but the $\{\theta_i\}$ set
%implicitly assumes that the scalar perturbations produced in the primordial Universe are not 
%affected by LQC for wavenumbers greater than the Hubble wavenumber today. This is a 
%fair hypothesis as otherwise the model would already be excluded by T- and E-mode measurements. 
%The "theoretical" uncertainties on $C^{B,prim}_\ell$ due to the 
%observationally-allowed small dispersion of parameters such as {\it e.g.} $\Omega_{CDM}$ will 
%then be much smaller  than the experimental uncertaintes considered hereafter. Those 
%"theoretical" errors can only marginally change our conclusions.

Although $k_\star,~R,~n_\mathrm{T}$ and $T/S $ can be translated into fundamental LQC 
parameters 
and specific initial conditions, we first leave them free as "generic phenomenological
parameters" so that they can be used to study different bouncing scenarios  (see, {\it e.g.}, 
\cite{peter} for a recent "classical bounce" investigation). Even if it was explicitly derived
in a LQC framework our parametrization is indeed quite general.

\subsection{Fisher analysis} 
In this framework, the question of a potential detection of the bounce in the 
B-mode anisotropies translates into specific values for $R$ and $k_\star$. 
To forecast the errors on the determination of those two parameters, we used 
a Fisher analysis method, as described in Ref. \cite{fisher}. (See also Ref. \cite{stivoli} for a more elaborated approach.) %Though more elaborated techniques incorporating foregournd subtraction in the error budget have been recently proposed \cite{stivoli}, the following Fisher analysis is sufficient for our purpose.
The $(5\times5)$ Fisher matrix reads
\begin{equation}
F_{ij}=\frac{1}{2}\displaystyle\sum_{\ell}\frac{1}{\Delta^2_\ell}\left.\frac{\partial 
C^B_\ell}{\partial\theta_i}\right|_{\theta_i=\bar\theta_i}\times\left.\frac{\partial 
C^B_\ell}{\partial\theta_i}\right|_{\theta_i=\bar\theta_j},
\end{equation}
where $C^B_\ell=C^{B,prim}_\ell+C^{B,lens}_\ell$ stands for the \{primordial+lensing\} B-mode spectrum and 
$\Delta_\ell$ is the error on the B-mode power spectrum recovery. We 
consider only the sampling and noise variance, {\it i.e.} 
$$\Delta^2_\ell=\frac{2}{(2\ell+1)f_{\text{sky}}}\left(C^B_\ell+\frac{N_\ell}{B^2_\ell}\right)^2,$$ 
where $B^2_\ell$ and $N_\ell$ are the power spectra of the Gaussian beam and 
the instrumental noise of the experiment, respectively, and $f_{\text{sky}}$ is the fraction of the sky 
used in the analysis. For a CMBPol/B-Pol-like mission, we relied on the experimental specifications of 
{\it EPIC-2m} \cite{epic} with an 8~arcmin beam, a noise level of 2.2~$\mu$K-arcmin, 
and a foreground separation accurate enough for a CMB power spectrum estimation 
using 70\% of the sky. 

To investigate the influence of degeneracies between parameters, the signal-to-noise 
ratio (SNR) for the $\theta_i$ parameters is computed in three different ways, performing
partial marginalization. We first assume a complete {\it ignorance} of the other four 
parameters, which results in $\mathrm{SNR}=\theta_i/\sqrt{[F^{-1}]_{ii}}$. Then we assume 
a perfect {\it knowledge} of the other parameters leading to 
$\mathrm{SNR}=\theta_i/\sqrt{[F_{ii}]^{-1}}$. Finally, we consider that only one 
parameter is known. If it is the $j$-th one, this translates to
$\mathrm{SNR}=\theta_i/\sqrt{[\mathcal{F}^{-1}]_{ii}}$ with $\mathcal{F}$ the 
$(4\times4)$ block of the Fisher matrix built by discarding the $j$-th raw and column. 
We finally search for the values of $\theta_i\equiv k_\star$ and $R$ such that SNR $>1(3)$ 
to define the $1\sigma(3\sigma)$ detectable values of these two parameters.

\section{Detecting the bounce: phenomenological parameters}
\label{sec:pheno}
\subsection{Detecting the transition length scale $k_\star$} 
The value of $k_\star$ is first varied from $10^{-6}$ to $1$ Mpc$^{-1}$. 
The fiducial values for the other four parameters are 
$\{R,~n_\mathrm{T},~T/S,~\tau\}=\{100,~0,~0.05,~0.087\}$ from which four classes of
models are generated by varying the parameters one by one: 
\begin{itemize}
\item class A: $R\in[10,10~000]$; 
\item class B: $n_\mathrm{T}\in[-0.1,0]$;
\item class C: $T/S\in[10^{-4},10^{-1}]$;
\item class D: $\tau\in[0,0.15]$. 
\end{itemize}

\begin{figure}
\begin{center}
\includegraphics[scale=0.48]{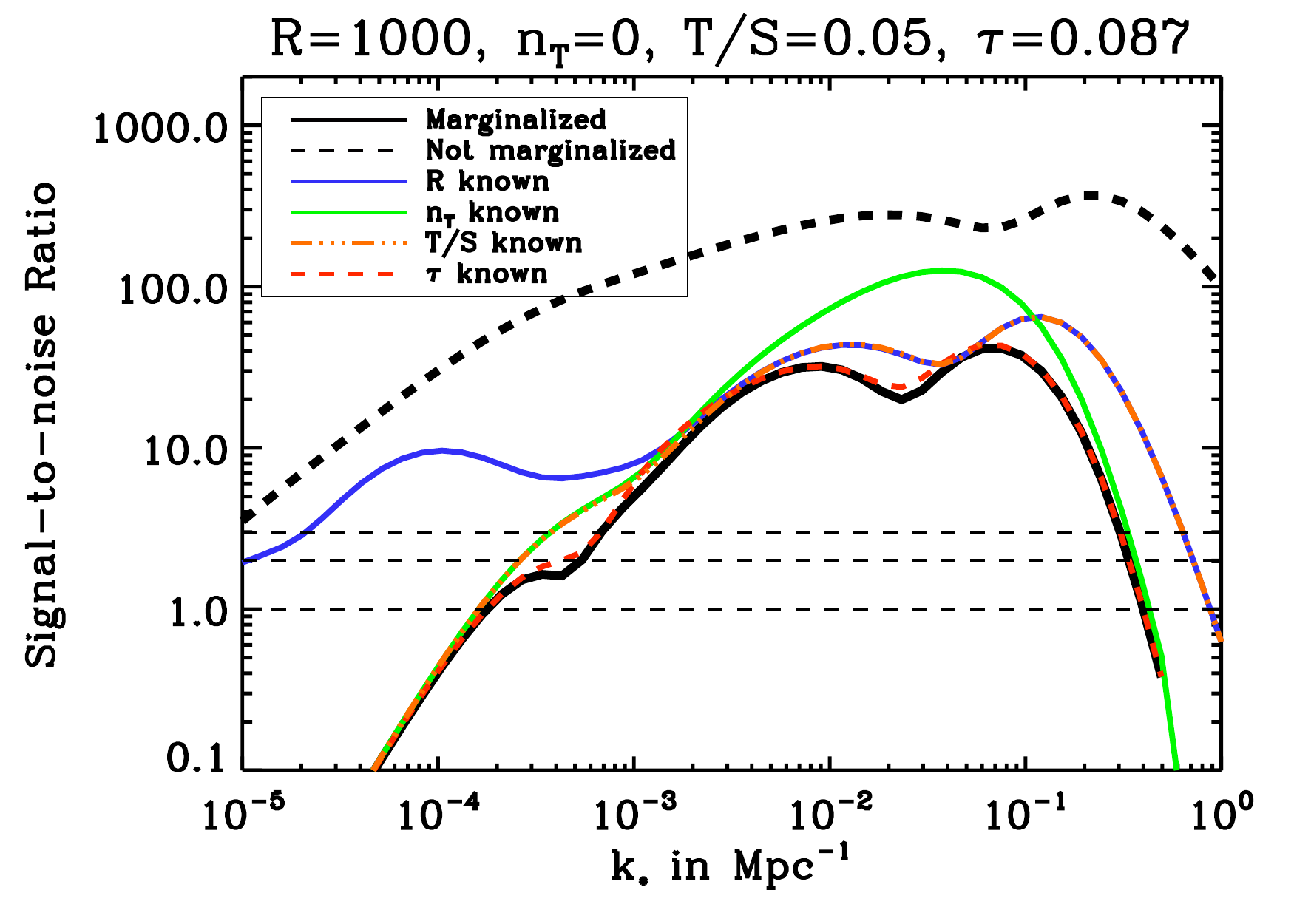}
\caption{SNR for $k_\star$ by performing partial marginalization. This shows that 
$k_\star$ is mainly degenerate with $R$ except for the tiny range 
$k_\star\sim10^{-2}-10^{-1}$~Mpc${}^{-1}$ where the main degeneracy is with the tensor spectral index. Horizontal lines stand for $1\sigma,~2\sigma$ 
and $3\sigma$ detection (from bottom to top).}
\label{snrk}
\end{center}
\end{figure}

As shown in Fig.~\ref{snrk}, which displays the SNR for 
$k_\star$ under different partial marginalizations, $k_\star$ is poorly degenerate with 
$T/S$ and $\tau$. (The dashed horizontal lines stand for 
$1\sigma$-,~$2\sigma$- and $3\sigma$-detections.) However, it is strongly degenerate first with $R$ for $k_\star<10^{-3}$ 
Mpc$^{-1}$, second with $n_\mathrm{T}$ for $10^{-3}<k_\star<10^{-1}$ Mpc$^{-1}$ and, third, 
with $R$ and $T/S$ for $k_\star>10^{-1}$ Mpc$^{-1}$. (As demonstrated in the next section, 
$k_\star$ is equally degenerate with $T/S$ and $R$ for $k_\star>10^{-1}$ Mpc$^{-1}$ because $R$ 
and $T/S$ are strongly degenerate in this regime.) The $(k_\star,n_\mathrm{T})$-degeneracy does 
not affect the potential detection of $k_\star$ as the fully marginalized SNR is already 
greater than 3 in the range where this degeneracy is dominant. However, comparing the 
solid-black and solid-blue curves shows that the range of $1\sigma$-detectable values of 
$k_\star$ is enhanced from $[1.5\times10^{-4},3\times10^{-1}]$ Mpc${}^{-1}$ to 
$[3\times10^{-6},9\times10^{-1}]$ Mpc${}^{-1}$ if the $(k_\star,R)$-degeneracy is broken. As a 
consequence, breaking this degeneracy could greatly enhance the potential of detection.

\begin{figure}
\begin{center}
\includegraphics[scale=0.48]{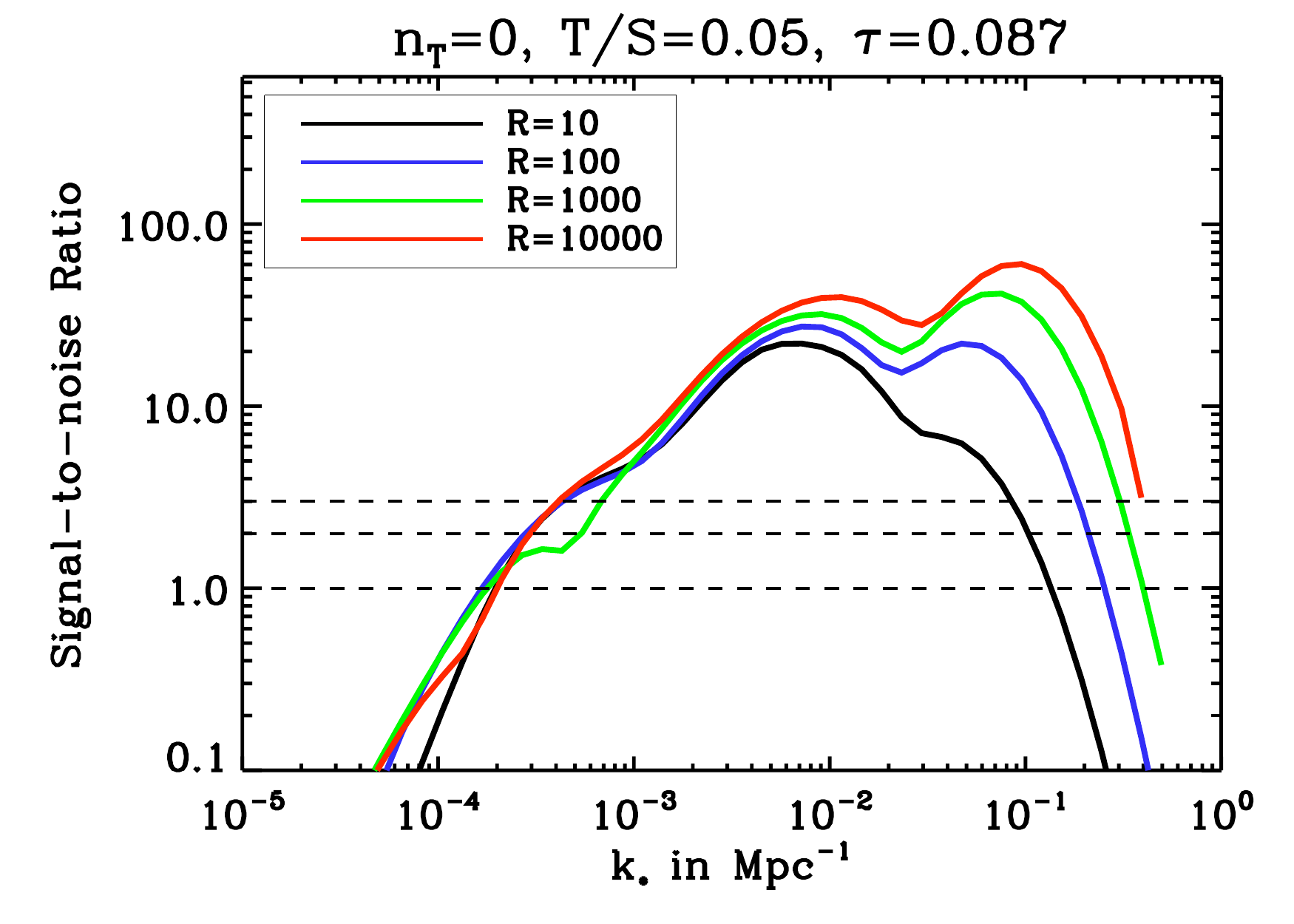}
\caption{Fully marginalized SNR for $k_\star$ with $R=10,~10^2,~10^3$ and $10^4$. (Horizontal lines are as in Fig.~\ref{snrk}.)}
\label{snrkresr}
\end{center}
\end{figure}

\begin{figure}
\begin{center}
\includegraphics[scale=0.48]{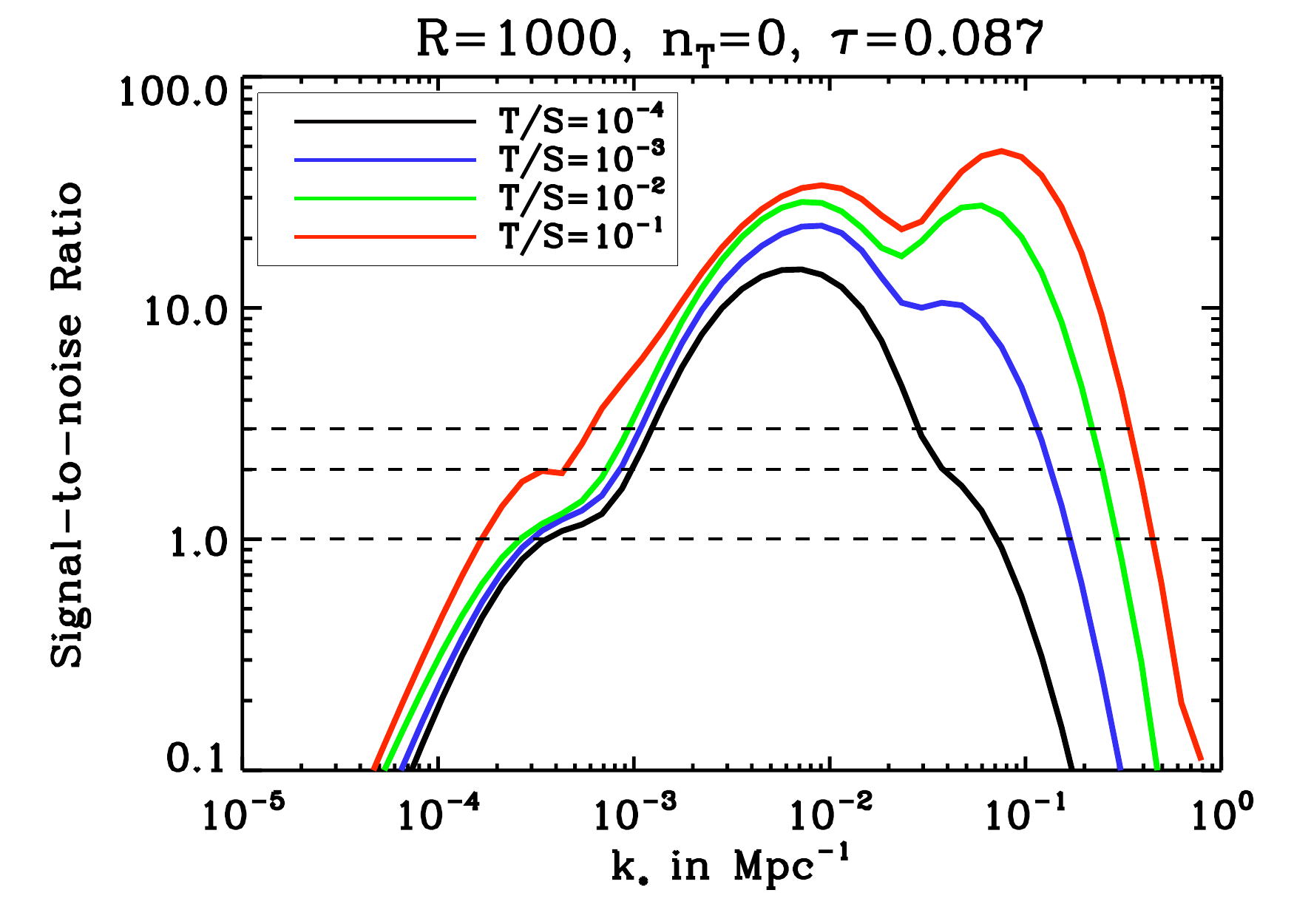}
\caption{Fully marginalized SNR for $k_\star$ with $T/S=10^{-4},~10^{-3},~10^{-2}$ and $10^{-1}$. (Horizontal lines are as in Fig.~\ref{snrk}.)}
\label{snrkrestsr}
\end{center}
\end{figure}

In Figs. \ref{snrkresr} and \ref{snrkrestsr}, the fully marginalized SNR for $k_\star$ is shown 
for four values of $R$ and four values of $T/S$ respectively. In both cases, this 
signal-to-noise ratio first increases with $k_\star$ as long as $k_\star<10^{-2}$ Mpc${}^{-1}$ 
and then decreases for higher values of $k_\star$. Higher values of $k_\star$ indeed
translate into a boost of the B-mode power for higher values of $\ell$ and the LQC distortion 
of $C^B_\ell$ is therefore located at multipoles with a smaller cosmic variance, explaining why 
the SNR first increases with $k_\star$. However, when $k_\star$ becomes greater than 
$\sim10^{-2}$ Mpc${}^{-1}$, the bump is shifted to $\ell>100$ and the B-mode power is 
strongly suppressed for $\ell<100$. As a consequence, for such high values of $k_\star$, the 
primordial part at large angular scales is hardly detectable because of its very faint power
and the boost at higher multipoles is completely masked by the lensing-induced B-mode, thus 
explaining why the SNR decreases for those higher values of $k_\star$.

\begin{table}
\begin{center}
	\begin{tabular}{cc||c|c} %\hline\hline
		\multicolumn{4}{c}{{\bf Full marginalization}} \\ \hline
		\multicolumn{2}{c||}{} & & \\[-6pt]
		\multicolumn{2}{c||}{model} & 1$\sigma$ & 3$\sigma$ \\ \hline
		& $R$ & & \\
		(A) & ${10^4}$ & ${[1.5\times10^{-4},6\times10^{-1}]}$   & ${[5\times10^{-4},4\times10^{-1}]}$  \\
		& 10 & ${[1.5\times10^{-4},1.5\times10^{-1}]}$   & ${[3\times10^{-4},8\times10^{-2}]}$   \\ \hline
		& $n_\mathrm{T}$ & & \\
		(B) & {0} & $[1.2\times10^{-4},3\times10^{-1}]$   & $[2.5\times10^{-4},2\times10^{-1}]$   \\
		& {-0.1} & $[1.2\times10^{-4},3\times10^{-1}]$   & $[2.5\times10^{-4},2\times10^{-1}]$  \\ \hline
		& $T/S$ & & \\
		(C) & ${10^{-1}}$ & $[1.2\times10^{-4},5\times10^{-1}]$   & $[6\times10^{-4},2.5\times10^{-1}]$ \\
		& ${10^{-4}}$ & $[3\times10^{-4},8\times10^{-2}]$   & $[1.2\times10^{-3},3\times10^{-2}]$  \\ \hline
		& $\tau$ & & \\
		(D) & {0.15} & $[1.2\times10^{-4},3\times10^{-1}]$   & $[3\times10^{-4},2\times10^{-1}]$ \\ 
		& {0.01} & $[2\times10^{-4},3\times10^{-1}]$   & $[4\times10^{-4},2\times10^{-1}]$ \\ \hline
		\multicolumn{4}{c}{} \\
		\multicolumn{4}{c}{{\bf No marginalization}} \\ \hline
		\multicolumn{2}{c||}{} & & \\[-6pt]
		\multicolumn{2}{c||}{model} & 1$\sigma$ & 3$\sigma$ \\ \hline
		& $R$ & &  \\
		(A) & ${10^4}$ & all range accessible  & ${[2\times10^{-6},1]}$   \\
		& 10 & ${[2\times10^{-5},1]}$   & ${[6\times10^{-5},7\times10^{-1}]}$   \\ \hline
		& $n_\mathrm{T}$ & & \\
		(B) & {0} & $[9\times10^{-6},1]$   & $[\times10^{-5},1]$   \\
		& {-0.1} & $[9\times10^{-6},1]$   & $[2\times10^{-5},1]$   \\ \hline
		& $T/S$ & & \\
		(C) & ${10^{-1}}$ & $[9\times10^{-6},1]$   & $[10^{-5},1]$   \\
		& ${10^{-4}}$ & $[9\times10^{-6},6\times10^{-1}]$   & $[2\times10^{-5},3\times10^{-1}]$   \\ \hline
		& $\tau$ & & \\
		(D) & {0.15} & $[10^{-5},1]$   & $[2\times10^{-5},1]$   \\ 
		& {0.01} & $[10^{-5},1]$   & $[2\times10^{-5},1]$   \\ \hline
	\end{tabular}
	\caption{Ranges of detectable values of $k_\star$ in Mpc$^{-1}$ by assuming complete ignorance (upper part) and perfect knowledge (lower part) of the other cosmological parameters.}
	\label{resultk}
\end{center}
\end{table}

\begin{figure*}
\begin{center}
	\includegraphics[scale=0.475]{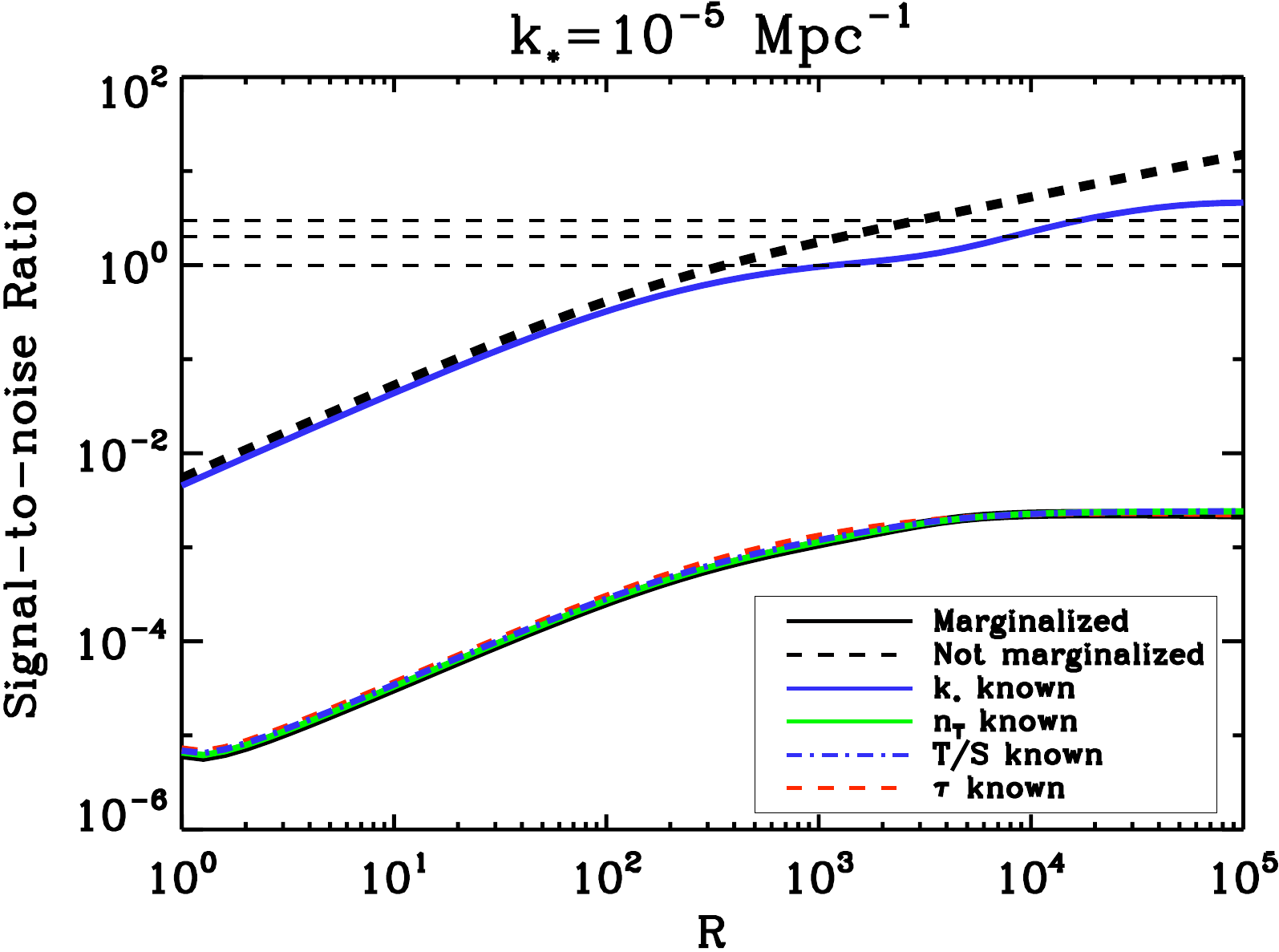} \includegraphics[scale=0.475]{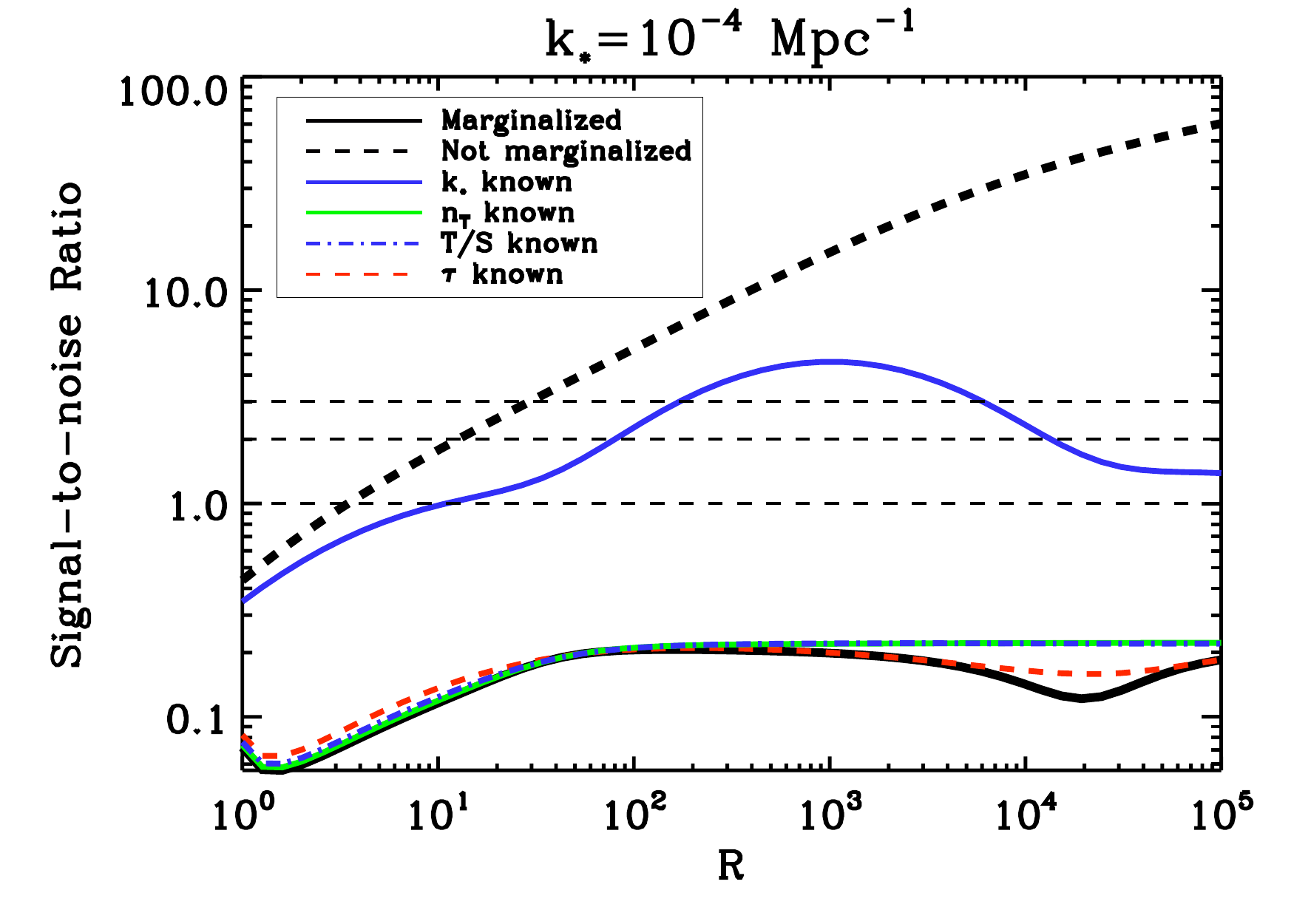} \\
	\includegraphics[scale=0.475]{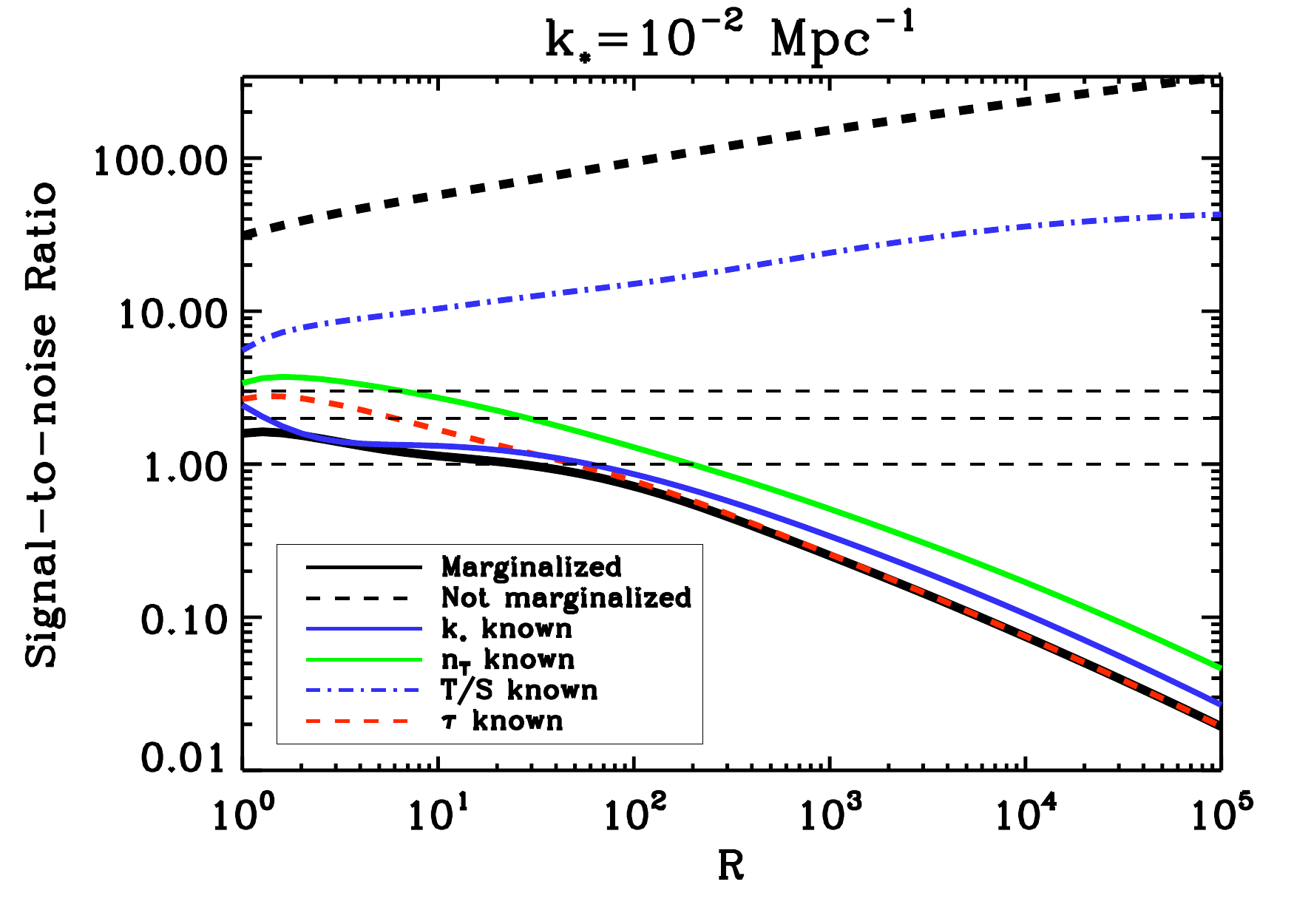} \includegraphics[scale=0.475]{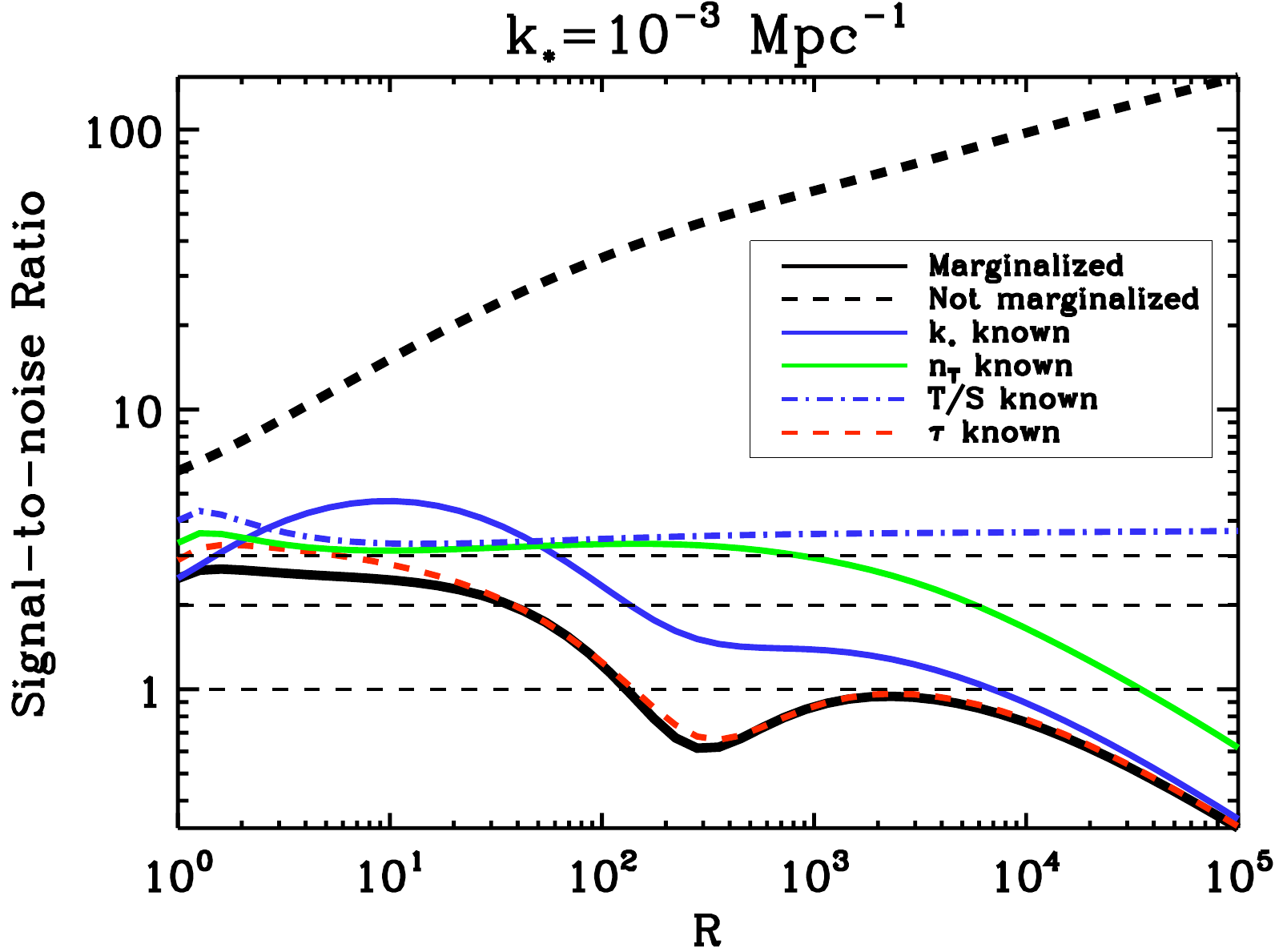} 
	\caption{SNR for $R$ in the class A model (see text) with different marginalization
	options. (Horizontal lines are as in Fig.~\ref{snrk}.)}
	\label{snrr-dege}
\end{center}
\end{figure*}

Moreover, our numerical investigations show that the shape of the SNR displayed in 
Figs. \ref{snrkresr} and \ref{snrkrestsr} is the same for all the considered models, 
which allows us to safely derive a range of
detectable values of $k_\star$. The $1\sigma$- and $3\sigma$-limits for a 
detection of $k_\star$ are given in Table \ref{resultk}. As one should have expected, the 
detection becomes possible for $k_\star\geq k_H$. Nevertheless, a detection of 
$k_\star<k_H$ is still possible as the tail of the bump may affect the B-mode power 
spectrum shape at large angular scales. As previously stated, this latter possibility clearly 
requires to break the $(k_\star,R)$-degeneracy. On the one hand, in the marginalized case, the minimum 
detectable value of $k_\star$ is affected by the values of $T/S$ and only very mildly depends  
on $R,~n_\mathrm{T}$ and $\tau$. On the other hand, the maximum detectable value of $k_\star$
depends on both $T/S$ and $R$ but does not depend on the specific values of $n_\mathrm{T}$ 
and $\tau$.

\subsection{Detecting the bump $R$}
Studying $R$ is more intricate as additional degeneracies have to be considered. 
Our fiducial model is given by 
$\{k_\star,~n_\mathrm{T},~T/S,~\tau\}=\{10^{-3}~\mathrm{Mpc}^{-1},~0,~0.05,~0.087\}$ 
and $R$ is varied from 1 to $10^5$. As for $k_\star$, we built four classes of models by 
varying each parameter: 
\begin{itemize}
\item class A: $k_\star~[\mathrm{Mpc}^{-1}]\in[10^{-5},10^{-2}]$; 
\item class B: $n_\mathrm{T}\in[-0.1,0]$; 
\item class C: $T/S\in[10^{-4},10^{-1}]$;
\item class D: $\tau\in[0,0.15]$.
\end{itemize}

In Fig.~\ref{snrr-dege}, $R$ is shown to be degenerate with different cosmological 
parameters. Depending on the value of $k_\star$, $R$ is either mainly degenerate 
with $k_\star$ (for low $k_\star$ values) or with  $T/S$ (for  high  $k_\star$ values). 
It was already clear from Fig.~\ref{snrk} that the $(k_\star,R)$-degeneracy is broken 
for $10^{-3}<k_\star<10^{-1}$ Mpc$^{-1}$. However, $R$ starts to be strongly
degenerate with $T/S$ for $k_\star>10^{-3}$ Mpc$^{-1}$. This explains
first why the marginalized SNR decreases for higher values of $R$ and second, why $k_\star$ 
appeared to be equally degenerate with $T/S$ and $R$ (see Fig. \ref{snrk}).
 The transition from 
the $(k_\star,R)$-degeneracy regime to the $(T/S,R)$-degeneracy regime occurs when 
$k_\star$ becomes close to the current Hubble scale. We stress that for 
$k_\star\sim10^{-3}$ Mpc$^{-1}$, the dichotomy between the $(k_\star,R)$- and 
$(T/S,R)$-degeneracy regimes is not meaningful as $R$ is here equally degenerate with 
$k_\star,~T/S$ and $n_\mathrm{T}$. Nevertheless, such an intricate situation corresponds to a 
very narrow range of $k_\star$ (see bottom-right panel of Fig. \ref{snrr-dege}).

\begin{table*}
\begin{tabular}{cc||c|c||c|c}
	%\multicolumn{6}{c}{} \\
	\multicolumn{6}{c}{$\mathbf{(k_\star,R)}${\bf -degeneracy regime $\mathbf{(k_\star<k_H)}$}} \\ \hline 
	& & \multicolumn{2}{c||}{} & \multicolumn{2}{c}{} \\[-6pt]
	& & \multicolumn{2}{c||}{$k_\star=10^{-5}$ Mpc$^{-1}$} & \multicolumn{2}{c}{$k_\star=10^{-4}$ Mpc$^{-1}$} \\[-6pt]
	& & \multicolumn{2}{c||}{} & \multicolumn{2}{c}{} \\
	& & Marg. & Not marg. & Marg. & Not marg. \\
	\multicolumn{2}{c||}{model} & $1\sigma~(3\sigma)$ & $1\sigma~(3\sigma)$ & $1\sigma~(3\sigma)$ & $1\sigma~(3\sigma)$ \\ \hline
	& & & & & \\[-6pt]
	(A) & & ${\mathrm{No~det.}}$ & $>3\times10^2~(>3\times10^3)$ & ${\mathrm{No~det.}}$ & ${>3~(>30)}$ \\ \hline
	& $n_\mathrm{T}$ & & & & \\
	(B) & 0 & ${\mathrm{No~det.}}$ & ${>200~(1200)}$ & ${\mathrm{No~det.}}$ & ${>1.3~(15)}$ \\
	& -0.1 & ${\mathrm{No~det.}}$ & ${>300~(3000)}$ & ${\mathrm{No~det.}}$ & ${>2~(30)}$ \\ \hline
	& $T/S$ & & & & \\ 
	(C) & ${10^{-1}}$ & ${\mathrm{No~det.}}$ & ${>400~(1000)}$ & ${\mathrm{No~det.}}$ & ${>3.5~(30)}$ \\  
	& ${10^{-4}}$ & ${\mathrm{No~det.}}$ & ${>1100~(10^5)}$ & ${\mathrm{No~det.}}$ & ${>10~(700)}$ \\ \hline
	& $\tau$ & & & & \\
	(D) & 0.15 & ${\mathrm{No~det.}}$ & ${>200~(1500)}$ & ${\mathrm{No~det.}}$ & ${>2~(15)}$ \\  
	& 0 & ${\mathrm{No~det.}}$ & ${>200~(1500)}$ & ${\mathrm{No~det.}}$ & ${>2~(15)}$ \\ \hline
	\multicolumn{6}{c}{} \\
	\multicolumn{6}{c}{$\mathbf{(T/S,R)}${\bf -degeneracy regime $\mathbf{(k_\star>k_H)}$}} \\ \hline
	& & \multicolumn{2}{c||}{} & \multicolumn{2}{c}{} \\[-6pt]
	& & \multicolumn{2}{c||}{$k_\star=10^{-3}$ Mpc$^{-1}$} & \multicolumn{2}{c}{$k_\star=10^{-2}$ Mpc$^{-1}$} \\[-6pt]
	& & \multicolumn{2}{c||}{} & \multicolumn{2}{c}{} \\
	& & Marg. & Not marg. & Marg. & Not marg. \\
	\multicolumn{2}{c||}{model} & $1\sigma~(3\sigma)$ & $1\sigma~(3\sigma)$ & $1\sigma~(3\sigma)$ & $1\sigma~(3\sigma)$ \\ \hline
	& & & & & \\[-6pt]
	(A) & & ${<100}$ ${\mathrm{(No~det.)}}$ & ${>1~\mathrm{at}~6\sigma}$ & ${<25}$ $\smmath{\mathrm{(No~det.)}}$ &  ${>1~\mathrm{at}~30\sigma}$ \\ \hline
	& $n_\mathrm{T}$ & & & & \\
	(B) & 0 & ${<200~\mathrm{and}~[600,10^4]~(30)}$ & ${>1~\mathrm{at}~8\sigma}$ & ${<100}$ ${\mathrm{(No~det.)}}$ & ${>1~\mathrm{at}~40\sigma}$  \\
	& -0.1 & ${<100}$ ${\mathrm{(No~det.)}}$ & ${>1~\mathrm{at}~6\sigma}$ & ${<8}$ ${\mathrm{(No~det.)}}$ & ${>1~\mathrm{at}~30\sigma}$ \\ \hline
	& $T/S$ & & & & \\ 
	(C) & ${10^{-1}}$ & ${<200~\mathrm{and}~[500,10^4]~(<2)}$ & ${>1~\mathrm{at}~6\sigma}$ & ${<100}$ ${\mathrm{(No~det.)}}$ & ${>1~\mathrm{at}~40\sigma}$ \\  
	& ${10^{-4}}$ & ${\mathrm{No~det.}}$ & ${>1.2~(20)}$ & ${\mathrm{No~det.}}$ & ${>6~(20)}$ \\ \hline
	& $\tau$ & & & & \\
	(D) & 0.15 & ${<10^4~(20)}$ & ${>1~\mathrm{at}~8\sigma}$ & ${<100~(3)}$ & ${>1~\mathrm{at}~40\sigma}$ \\  
	& 0 & ${<150~\mathrm{and}~[300,22000]~(<60)}$ & ${>1~\mathrm{at}~8\sigma}$ & ${<2~\mathrm{and}~[4,100]}$ ${\mathrm{(No~det.)}}$ & ${>1~\mathrm{at}~40\sigma}$ \\ \hline
\end{tabular}
\caption{Range of detectable values of $R$. Because of degeneracies, this range mainly depends on $T/S$ and $k_\star$.} %In the $(k_\star,R)$-degeneracy regime (upper part), a minimal detectable values of $R$ is determined while in the $(T/S,R)$-degeneracy regime, a maximal detectable value of $R$ is found.}
\label{resultr}
\end{table*}

Because of this $k_\star$-dependent degeneracy, {\it meaningful} results concerning the detection of $R$ also necessarily depend on $k_\star$. For each class of model, we provide results for $k_\star=10^{-5},~10^{-4},~10^{-3}$ and $10^{-2}$ Mpc$^{-1}$, as summarized in Table \ref{resultr}. If degeneracies are indeed broken (i.e., no marginalization over  $\{k_\star,~n_\mathrm{T},~T/S,~\tau\}$), the SNR increases for higher values of $R$. This remains true if marginalization is performed for $k_\star\leq k_H$ (i.e., in the $(k_\star,R)$-degeneracy regime), allowing us to derive a lowest detectable value of $R$. In the remaining cases (i.e., in the $(T/S,R)$-degeneracy regime), the SNR {\it decreases} for higher values of $R$, leading to upper limits on $R$. As can be concluded from Table \ref{resultr}, a detection of $R$ requires one to break the degeneracies if $k_\star\leq k_H$ while a detection up to a couple of thousands is possible {\it without} breaking the degeneracies if $k_\star\geq k_H$.
	
\section{Detecting the bounce: cosmological interpretation}
\label{sec:fund}
\subsection{Fundamental parameters of the LQC-universe} 
Let us now translate those constraints into constraints on the
fundamental parameters of LQC.  Interestingly, it can be shown that the fundamental parameters (describing either the field
itself, the initial conditions or the LQC corrections) are quite simply related with the 
observable parameters previously defined, {\it i.e.} $k_\star,~R,~n_\mathrm{T}$ and $T/S$. To derive the following relations, we took into account the LQC corrections for the background
dynamics (which leads to the bounce) and for the propagation of 
gravitational waves \cite{jakub2010b}.

First of all, the bump amplitude 
is well approximated by $R \approx (m_{\text{Pl}}/m_\phi)^{0.64}$ (see our detailed analysis
 presented in \cite{jakub2010b}) with $m_\text{Pl}=1.22\times10^{19}$~GeV the Planck mass and $m_\phi$ the inflaton mass, {\it i.e.}, 
 $$
 V(\phi)=\frac{1}{2}m^2_\phi\phi^2.
$$

Second, by computing the expansion of the universe since the time when 
$k_{\star}$ crossed the horizon and rewriting the different terms 
entering this ratio (in particular, the number of e-folds during inflation
being given by 
$N_{\text{inf}}\approx(4\pi/m^2_{\text{Pl}})\int_0^{\phi_{\text{max}}}(V/V')d\phi$), 
one can show that the transition scale $k_\star$ is given by
\begin{equation}
k_\star=\frac{\frac{4\pi^{\frac{3}{2}}}{\sqrt{3}}\frac{m_\phi}{m_{\text{Pl}}}\phi_{\text{max}}}{\exp\left(
2\pi\frac{\phi_{\text{max}}^2}{m_{\text{Pl}}^2}\right) \frac{T_{\text{RH}}}{T_{\text{eq}}}
\left( \frac{g_{\text{RH}}}{g_{\text{eq}}} \right)^{\frac{1}{3}}(1+z_{\text{eq}})}, \label{kstar}
\end{equation}
where $\phi_{\text{max}}$ is the maximum value of the field, $m_\phi$ is
its mass, $T_{\text{RH}}$ and $g_{\text{RH}}$ are the reheating 
temperature and the corresponding number of degrees of freedom, respectively, and  
$T_{\text{eq}}\simeq0.75$~eV, $z_{\text{eq}}\simeq3196$ and $g_{\text{eq}}\simeq3.9$ 
are the temperature, redshift and degrees of freedom at matter/radiation equality, respectively (see {\it e.g.},~Sec. 3.4.4 of Ref. \cite{wmap5}).
In addition, numerical investigations have shown that $\phi_{\text{max}}$ 
can be straightforwardly related with the "initial conditions" or, 
more precisely, with the physical conditions at the bounce: 
\begin{equation}
\phi_{\text{max}}\approx \phi_{\text{bounce}}+m_{\text{Pl}} = \left(\frac{\sqrt{2\rho_c}}{m_\phi}\right)x+m_{\text{Pl}}.
\label{phimax}
\end{equation}
In this expression,  $\phi_{\text{bounce}}$, $\rho_c$ and $x^2=V(\phi_{\text{bounce}})/\rho_c$ correspond, respectively, to the value of the scalar field, the total energy density and the
fraction of potential energy {\it at the bounce}. The value of the total energy density at the 
bounce could be considered as a free parameter of the theory. However, if the Barbero-Immirzi 
parameter is taken at the value required to recover the Bekenstein black hole entropy, 
{\it i.e.}, $\gamma\simeq0.239$, this leads to 
$\rho_c\approx0.82m^4_{\text{Pl}}$. The number of e-folds during 
inflation is  given by $\rho_c$ and by the ratio $x/m_\phi$, through
$$
N_{\text{inf}}\approx\frac{2\pi}{m^2_\text{Pl}}\left[\left(\frac{\sqrt{2\rho_c}}{m_\phi}\right)x+m_{\text{Pl}}\right]^2.
$$
For the above-given value of $\rho_c$, a minimum amount of 60 e-folds during inflation is 
achieved if $x\geq1.64m_\phi/m_{\text{Pl}}$.

It is worth noticing that the number of fundamental parameters is smaller than the number of 
phenomenological ones ($T_{\text{RH}}(g_{\text{RH}})^{1/3}$ acting as a unique effective 
parameter) which leads to a kind of consistency relations for the LQC parameters. Moreover, 
the $(k_\star,R,T/S)$-degeneracies being partially broken by restricting the cosmological 
interpretation to LQC, the detection of a LQC-induced bounce is {\it a priori} more likely 
than the general detection of a bounce. However, we adopt a conservative approach and keep 
track of the different degeneracies appearing at the phenomenological level by using the 
fully marginalized limits derived on $k_\star$ and $R$.

\subsection{Detecting fundamental parameters}

\subsubsection{Probing the model with future B-mode experiments}

As previously explained, the LQC corrections to scalar modes are
not yet known. As a first hypothesis, we therefore assume that the 
temperature spectrum (the one which is very well measured by 
WMAP and is about to be still improved by Planck) is not affected. 
In this case, nearly no constraint can be put with current data and 
the study is purely prospective. The question we want to answer is then the following: 
In which range
should the fundamental parameters lie for the LQC effects to be detected through the B-mode
spectrum modifications? The 
amplitude of the expected bump is set by the mass of the field and 
the value of the transition scale $k_\star$ is set by both the mass of the field and the 
initial conditions. From the observational 
viewpoint, $k_\star$ is by far the most important parameter. We will therefore translate the 
detectable range of $k_\star$ into detectable regions in the $(m_\phi,\phi_{\text{max}})$ and 
$(m_\phi,x)$ planes. \\

The first estimate can be very easily obtained. Basically, the IR suppression predicted by the
model becomes observable when $k_\star$ is high enough (otherwise, the suppression occurs
only on superhorizon scales). This translates into an upper limit on $\phi_{\text{max}}$ and
therefore into an upper limit on $x$. By assuming the usual $m_\phi\approx 10^{-6}$ value, the
numerical analysis
leads to $x<2\times10^{-6}$: The bounce can be discriminated from the standard prediction when $x$ is very small. 
It means that the LQC effects appear in the B-mode spectrum when the universe is strongly 
dominated by  kinetic energy at the bounce. This is a
consistent conclusion as backreaction effects should anyway be added when the potential energy
becomes important. \\

From the detection viewpoint, a more refined estimate can be obtained by using  the details of the previous analysis. In
this case we require not only that the features of the \{bouncing+inflationary\} model differ
from that standard prediction but also that they can be detected {\it by themselves}. This is
by far more constraining. In this case, the effects become observable when $k_\star$ 
lies within a restricted interval. For a fixed value of $m_\phi$, the lower(upper) bound
on $k_\star$ can still be translated into an upper(lower) limit on $\phi_{\text{max}}$ 
(except for a tiny parameter space corresponding to unrealistically 
small values of $\phi_{\text{max}}$) and therefore into an upper(lower) limit
on $x$. On the opposite, for a fixed value of $x$, the lower(upper) bound on $k_\star$ is 
translated into a lower(upper) limit on $m_\phi$. Translating "detectable $k_\star$" into 
"detectable $(m_\phi,\phi_{\text{max}},x)$" is however plagued by two types of uncertainties.
First of all, neither the reheating temperature nor the number of degrees of freedom are 
known. We will therefore let $T_\text{RH}$ vary from $10^{10}$ to $10^{16}$
and $g_\text{RH}$ vary from its standard model value to its supersymmetry value. 
Second, the detectable range of $k_\star$ depends on the values of the other cosmological 
parameters. From the fully marginalized $1\sigma$-detection presented in Table 
\ref{resultk}, we define three possible ranges of detectable values of $k_\star$:
\begin{itemize}
\item pessimistic: $\left[3\times10^{-4},8\times10^{-2}\right]$~Mpc${}^{-1}$,
\item intermediate: $\left[2\times10^{-4},3\times10^{-1}\right]$~Mpc${}^{-1}$, and
\item optimistic: $\left[1.5\times10^{-4},6\times10^{-1}\right]$~Mpc${}^{-1}$.
\end{itemize}
We stress out that this last uncertainty is mainly associated with the upper bound on 
$k_\star$. This means that the lower(upper) limit on $x(m_\phi)$ will be mainly affected by 
uncertainties on other cosmological parameters than the transition scale, especially $R$ and $T/S$. (We recall that the above defined detectable ranges account for the different degeneracies. In particular, this range is greatly broadened if the degeneracies with either $R$,  $T/S$, or both are broken. This considerably widens the achievable region of LQC-parameter space. However, as our "translation" is solely based on the potential detection of $k_\star$, conservative forecasts should incorporate our "ignorance'" of {\it e.g.}, $R$.)

\begin{figure*}
\begin{center}
\includegraphics[scale=0.48]{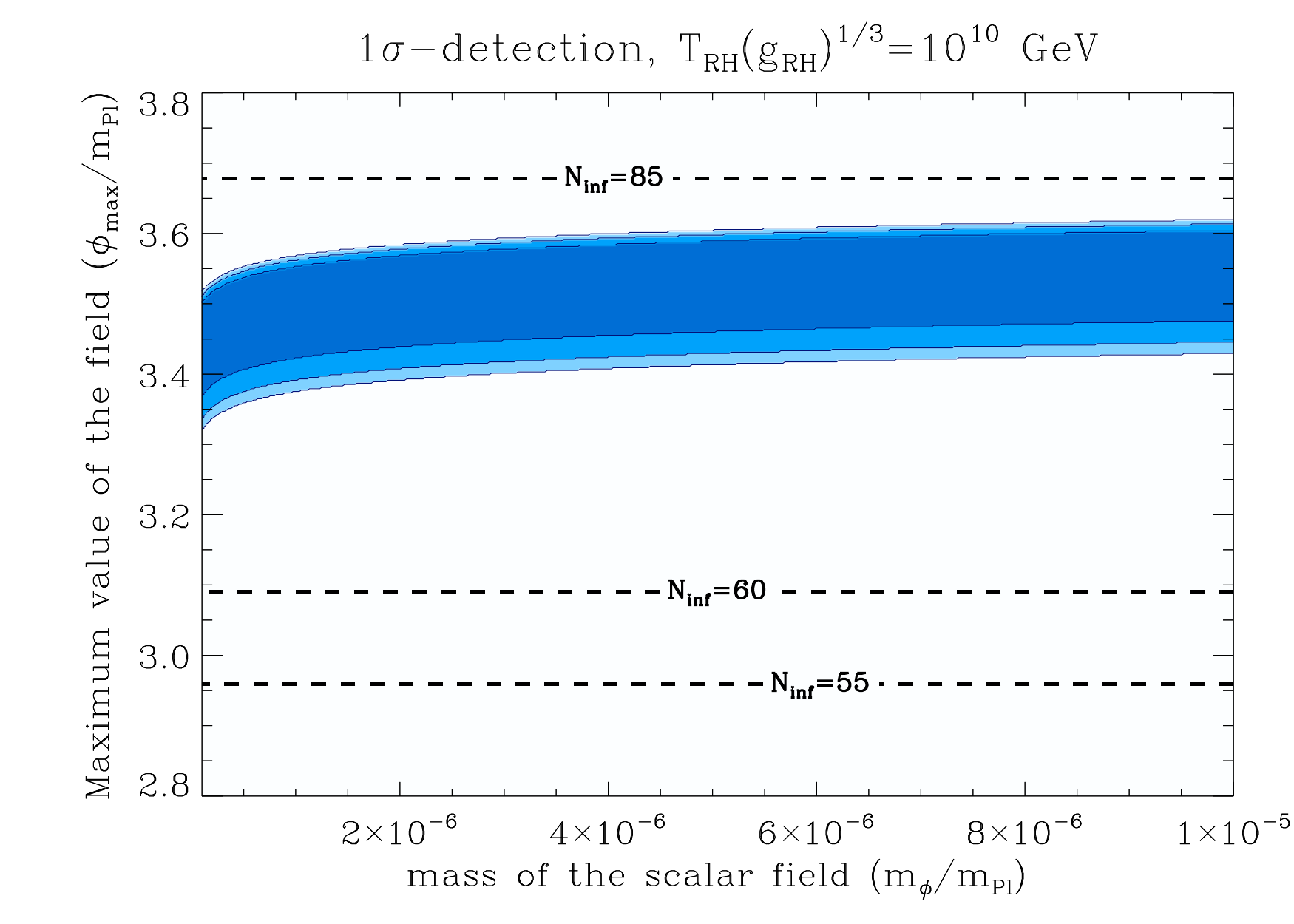} \includegraphics[scale=0.48]{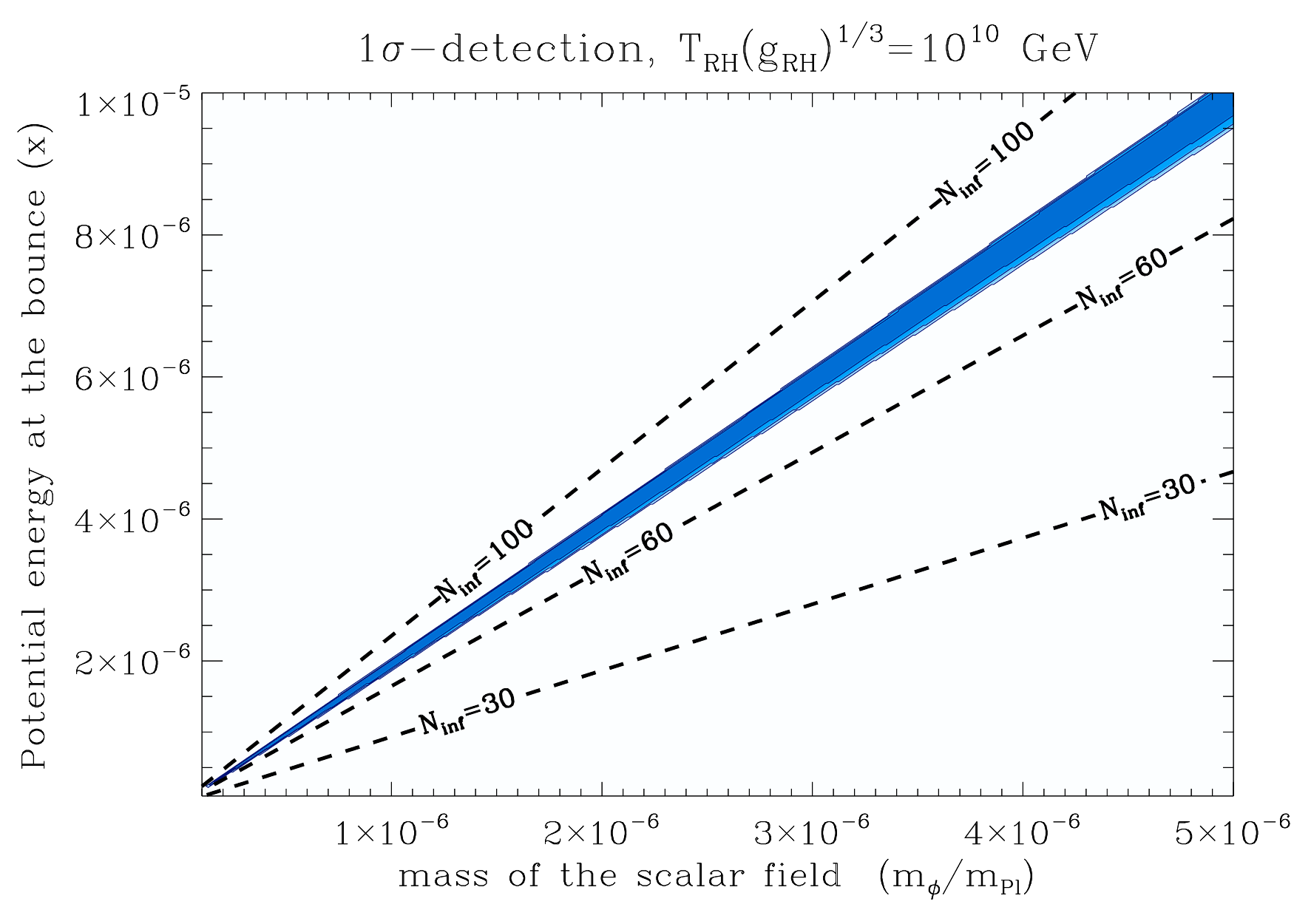} \\
\includegraphics[scale=0.48]{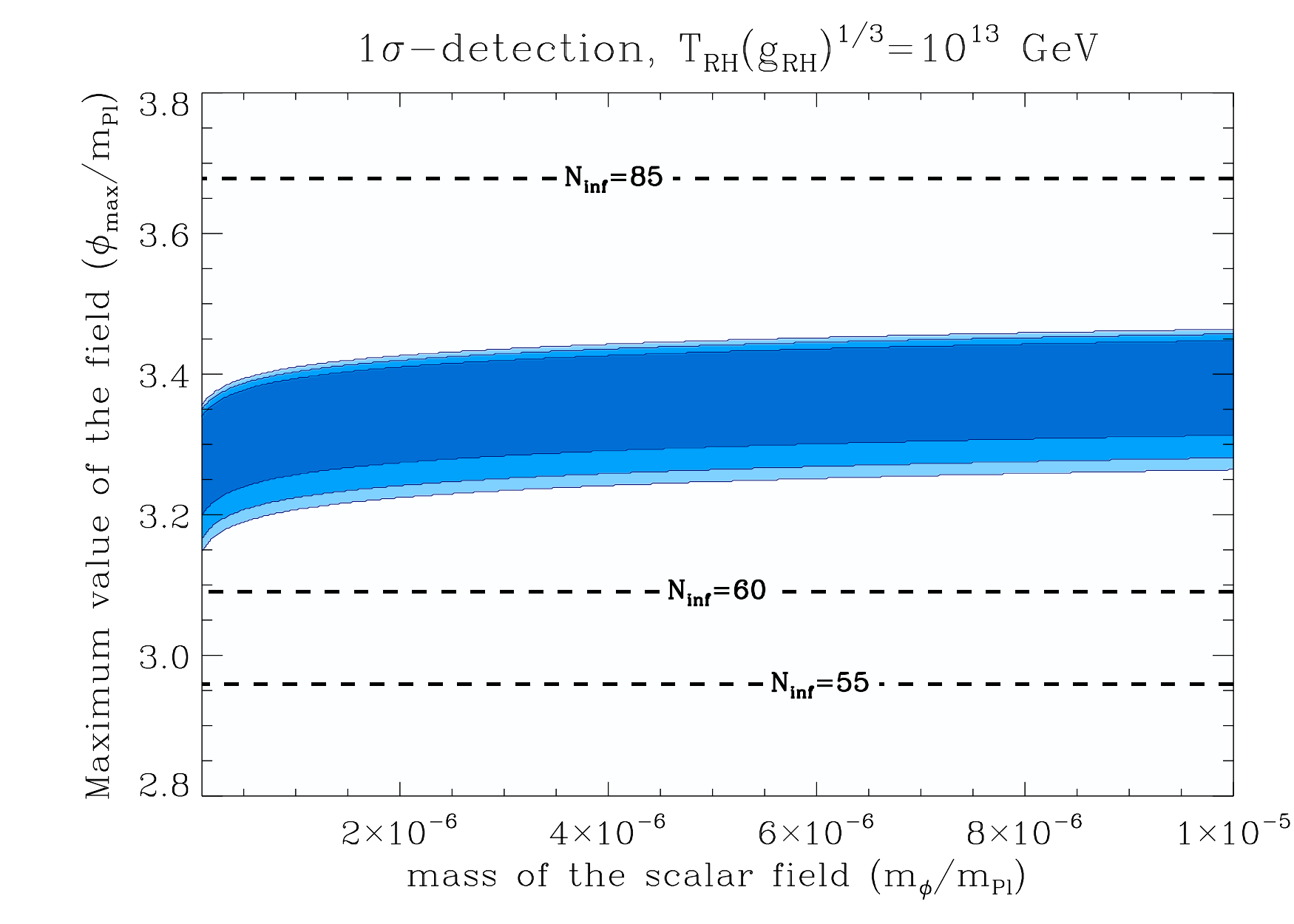} \includegraphics[scale=0.48]{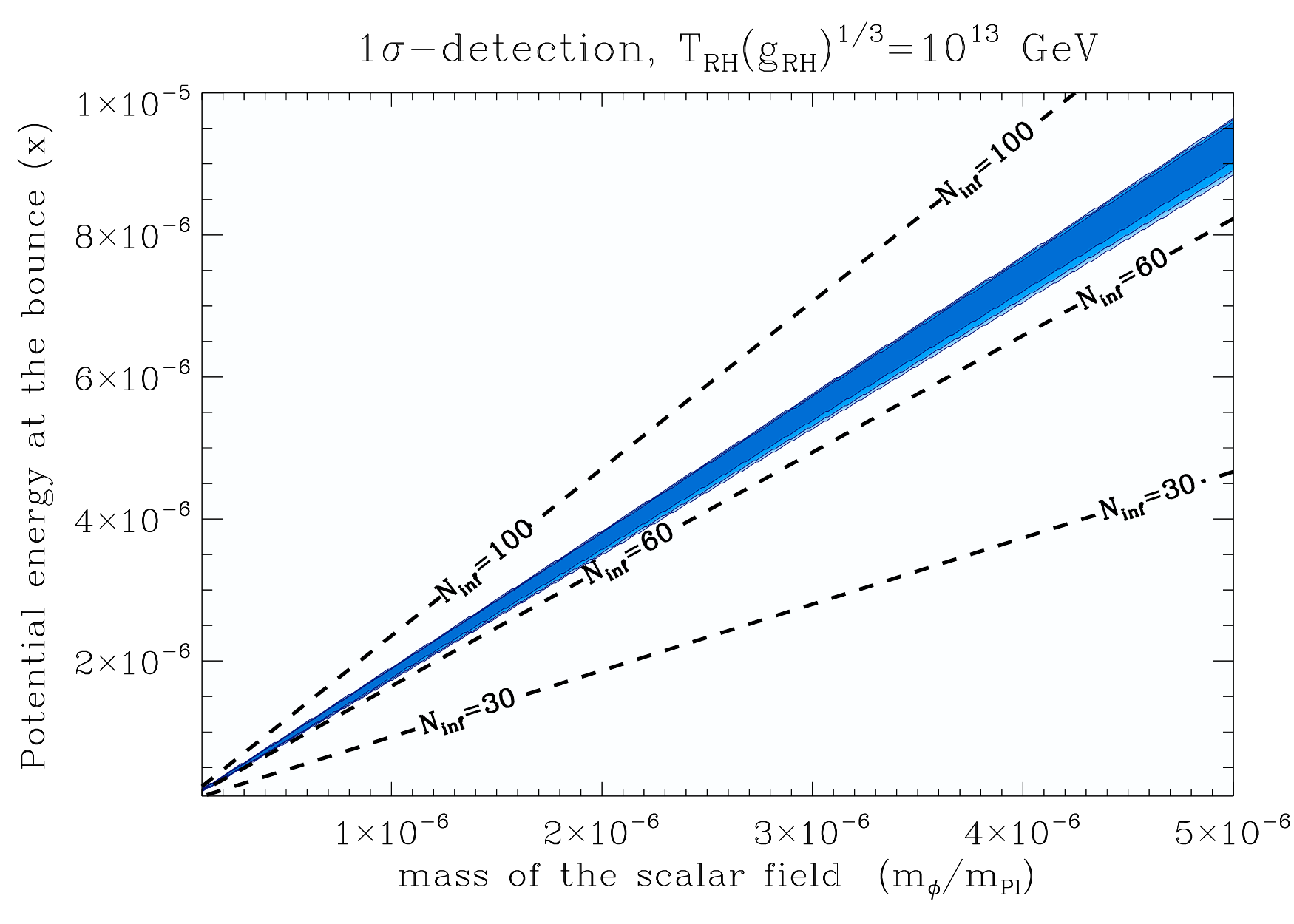}\\
\includegraphics[scale=0.48]{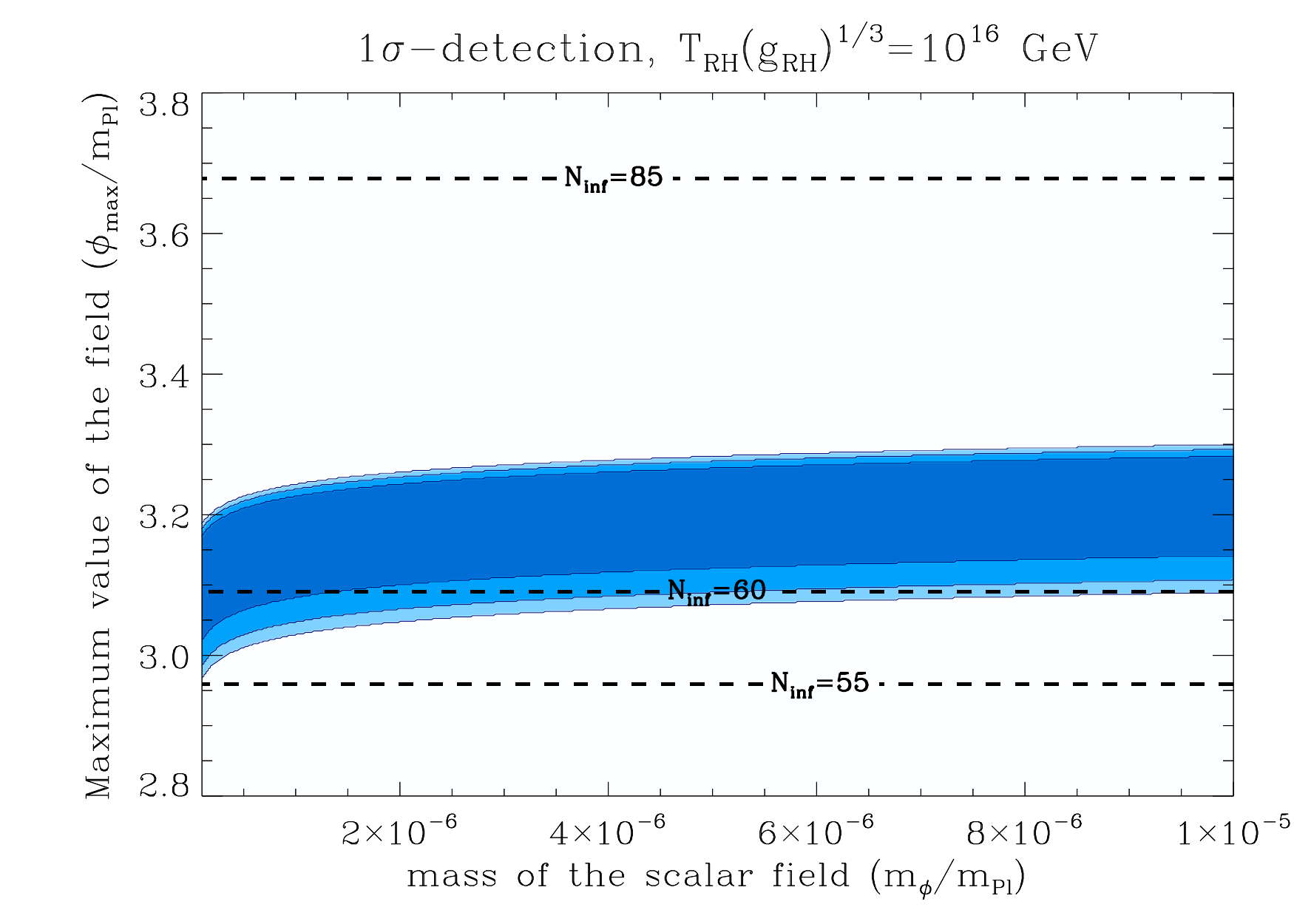} \includegraphics[scale=0.48]{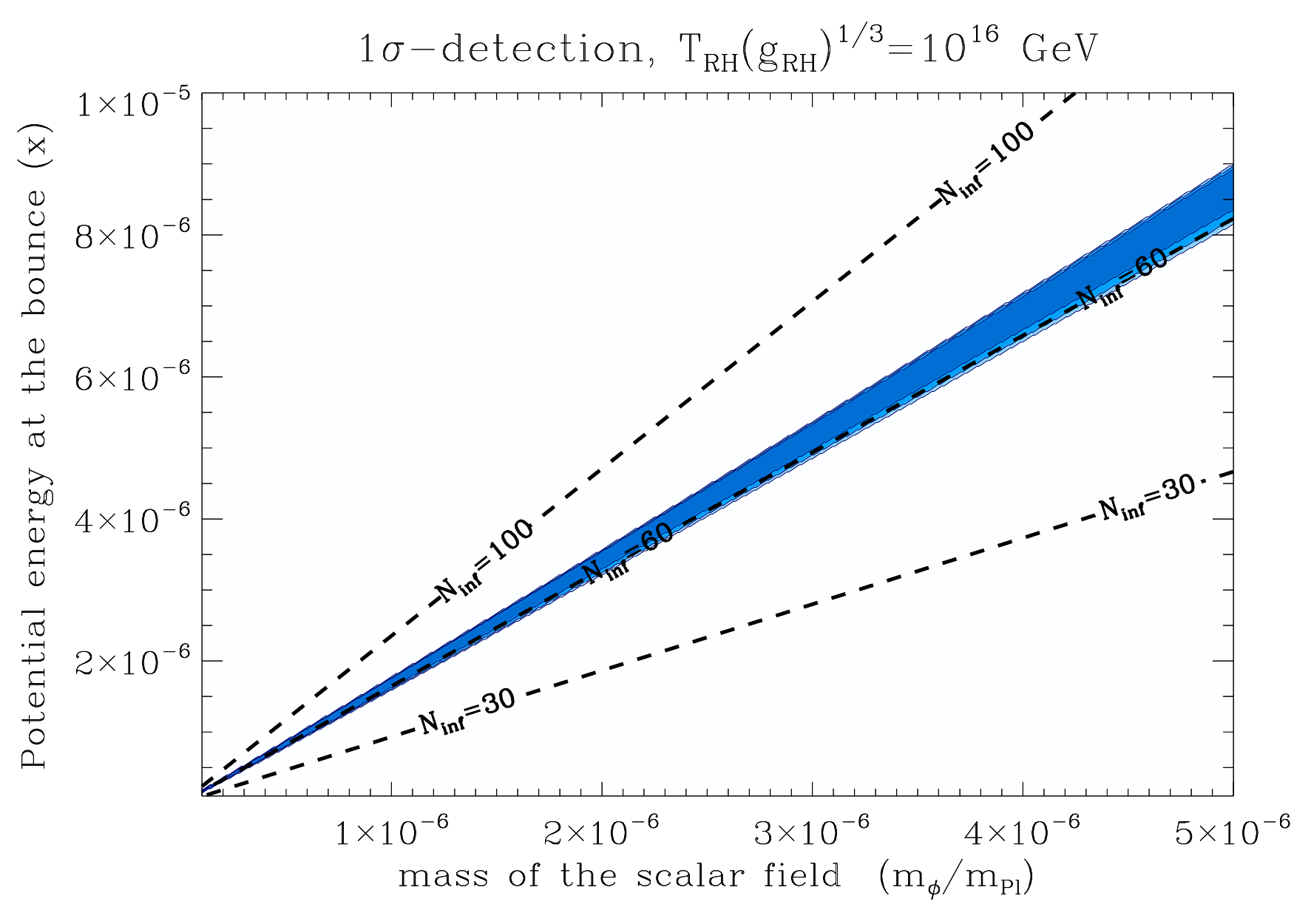}
\caption{1$\sigma$-detection of $(\phi_{\mathrm{max},}m_\phi)$ (left panel) and $(x,m_\phi)$ 
(right panel) as would be obtained from a detection of $k_\star$ in the B-mode power spectrum 
by assuming three different values of the "reheating parameter", {\it i.e.} 
$T_\mathrm{RH}\left(g_\mathrm{RH}\right)^{1/3}=10^{10},~10^{13}$, and $10^{16}$ GeV. 
Three ranges of detectable $k_\star$ are considered (see the core of the text) and lighter to darker blue runs from the 
most optimistic to the less optimistic case.}
\label{exclusion}
\end{center}
\end{figure*}

Our numerical results are summarized in Fig. \ref{exclusion}. It displays the detectable 
regions in the $(m_\phi,\phi_\text{max})$ and $(m_\phi,x)$ planes for 
different values of the reheating parameters, $T_\text{RH}\left(g_\text{RH}\right)^{1/3}=10^{10},~10^{13}$ and $10^{16}$ GeV. Lighter to 
darker blue goes from the most optimistic to the most pessimistic ranges of detectable 
$k_\star$.

The left panels of this figure clearly show that the results do not depend a lot
on the choice of the detectable $k_\star$ range.
The conclusions are therefore
robust with respect to changes of $R,~n_\mathrm{T},~T/S$ and $\tau$. 

The detection 
region for $(m_\phi,x)$ lies between two straight lines. Their slopes are fixed, first, 
by $T_\text{eq},~g_\text{eq},~z_\text{eq},T_{\text{RH}}(g_\text{RH})^{1/3}$ and, second, 
by $k_{\star,\text{max}}$ for the lower line and $k_{\star,\text{min}}$ for the upper line, with $k_{\star,\text{max(min)}}$ the upper(lower) bound of the detectable values of $k_\star$. 
On defining 
$$
\Delta_\text{RH}=\frac{4\pi^{3/2}}{\frac{T_{\text{RH}}}{T_{\text{eq}}}
\left( \frac{g_{\text{RH}}}{g_{\text{eq}}} \right)^{\frac{1}{3}}(1+z_{\text{eq}})\sqrt{3}},
$$
the transition scale is recast as a function of $m_\phi$ and $x$ as follows:
$$
k_\star=\frac{\Delta_\text{RH}\left[\left(\frac{\sqrt{2\rho_c}}{m_\text{Pl}}\right)x+m_\phi\right]}{\exp\left[\frac{2\pi}{m^2_\text{Pl}}\left(\left(\frac{\sqrt{2\rho_c}}{m_\phi}\right)x+m_{\text{Pl}}\right)^2\right]}.
$$
Except in a very narrow range, a variation of either $x$ or $m_\phi$ would mostly influence
$k_\star$ via the exponential. We can therefore approximate the numerator by a constant, 
dubbed $\mu_{x,m_\phi}$, to get
\begin{equation}
x\left(\frac{m_\text{Pl}}{m_\phi}\right)\approx\frac{m^2_\text{Pl}}{\sqrt{2\rho_c}}\left[\sqrt{\frac{\ln\left(\mu_{x,m_\phi}\Delta_{\text{RH}}/k_\star\right)}{2\pi}}-1\right]
\label{eq:last}
\end{equation}
As $\Delta_\text{RH}$ decreases for higher values of $T_{\text{RH}}(g_\text{RH})^{1/3}$, this
roughly explains why the slope of the detectable region in the $(m_\phi,x)$ plane shifts down 
for higher reheating temperatures. Moreover, the logarithmic dependence of this detectable 
region with $T_{\text{RH}}(g_\text{RH})^{1/3}$ underlines the robustness of our results. 

Finally,  a detection of $k_\star$ essentially constrains the values of the ratio $(x/m_\phi)$ explaining why a wide band in the $(m_\phi,x)$ plane is {\it a priori} detectable, including large values of $x$ and $m_\phi$. Nevertheless, the fact that arbitrary small values of $x$ can be detected means, once again,
that the LQC effects appear when the universe is strongly dominated by 
kinetic energy at the bounce. (Moreover, and as explained before, $m_\phi\sim10^{-6}~m_\text{Pl}$ being favored, this translates into a detectable value of $x\sim10^{-6}$.)  \\

Let us summarize our results. Calling $\alpha(\Delta_\text{RH},k_\star)$ the right-hand side of Eq. (\ref{eq:last}), a {\it detection} of the LQC-induced bounce is obtained if 
$$
x\left(\frac{m_\text{Pl}}{m_\phi}\right)\in[\alpha(\Delta_\text{RH},k_{\star,\text{max}}),\alpha(\Delta_\text{RH},k_{\star,\text{min}})].
$$
However, {\it discriminating} between the standard inflationary prediction and the LQC prediction requires only that
$$
x\left(\frac{m_\text{Pl}}{m_\phi}\right)\leq\alpha(\Delta_\text{RH},k_{\star,\text{min}}).
$$
(We recall here that higher values of $k_\star$ lead to smaller values of $\alpha$.)

It should also be pointed out that $k_\star$ 
can also be directly related to $\phi_{\text{obs}}$ (the value of the field
when the pivot mode crossed the horizon), which is itself related with 
the tilt of the scalar spectrum \cite{jakub2010c}. The results based on this method are basically the same.\\

\subsubsection{Constraining the parameters with available data}

Most of the corrections to the spectrum are {\it not} 
due to subtle LQC effects on the propagation of physical modes but to 
the bounce in itself. Unless some quite surprising cancellation occurs, 
it is therefore reasonable to assume that scalar modes are in fact modified 
in a quite similar way. Under this assumption, one can already use the current
data to constrain the model. As no $k^2$ infrared suppression is observed 
in the scalar power spectrum, it means that $x>2\times10^{-6}$. Stated otherwise, 
most of the parameter space of the theory is in agreement with the data. This is important
as it was  demonstrated that most of the parameter space also leads
to a long enough inflation phase (with more than 60 e-folds; see \cite{ashtekar2010}).
%In addition, one can conjecture that although higher than $2\times10^{-6}$, $x$ is not {\it much} higher than this, otherwise the slow-roll conditions would be violated for $\phi\approx\phi_{\text{obs}}\ll\phi_{\text{max}}$. This leaves open the window for detection.\\

\section{Conclusion}
\label{sec:concl}
%\paragraph*{Conclusion.} 
In this article, we have carefully investigated how next-generation B-mode 
CMB experiments could probe big bounce footprints. Under very general 
assumptions, it was demonstrated that, as far as  phenomenological parameters are concerned,
a substantial parameter space 
could be investigated. Furthermore, it was pointed out that this also makes 
quantum gravity effects possibly observable, especially in the LQC framework.

\section*{Remarks}
%Recently, a similar and independent study has been released \cite{ma}. It relies on the
%use of $k_\star $ and $m_\phi$ as 
%{\it cosmological parameters} and can be viewed as a kind of "mixing" of the {\it phenomenological} 
%and {\it fundamental} approaches here developped. Moreover, the number of e-folds is set to a fixed value of 
%60. This turns out to break the $(T/S,R)$-degeneracy  --those two phenomenological 
%parameters being both unambiguously determined by $m_\phi$ only-- and one should therefore 
%consider our analysis as more conservative.

Recently, two similar and independent studies have been released in \cite{ma} and \cite{liu}. The former \cite{ma} relies on Eq. (\ref{eff1}) to parametrize the primordial tensor power spectrum. It
uses $k_\star $ and $m_\phi$ as 
{\it cosmological parameters} and can be viewed as a kind of "mixing" of the {\it phenomenological} 
and {\it fundamental} approaches here developed. Moreover, the number of e-folds is set to a fixed value of 
60. This turns out to break the $(T/S,R)$-degeneracy  --those two phenomenological 
parameters being both unambiguously determined by $m_\phi$ only-- and one should therefore 
consider our analysis as more conservative. The latter \cite{liu} is based on scalar perturbations with a parametrization involving a jump at a given transition wavenumber but no additional bump. From a joint likelihood analysis on temperature CMB power spectrum from WMAP 7-yr, matter power spectrum from SDSS and SNIa "Union" compilation, a similar upper limit on the transition scale of the order of $2.44\times10^{-4}$~Mpc$^{-1}$ has been derived.

Second, the cosmological interpretation in terms of a bouncing universe induced by LQC obviously depends on the robustness of the underlying model. In particular, such a bouncing scenario is achieved by considering a homogeneous universe only and the bounce may not survive in models incorporating inhomogeneous degrees of freedom. Such an open question is still debated. Nevertheless, the above presented study remains relevant for two reasons: First of all, the phenomenological results displayed in Sec. \ref{sec:pheno}, though apparently motivated by LQC, apply to any models predicting a tensor power spectrum with a shape identical to the here-assumed one. Second, though previous studies pinned down that the bounce may not survive to inhomogeneities \cite{thiemann}, some recent studies based on the dipole approximation of LQG suggest the opposite \cite{battisti}.

\acknowledgments
Some of the results in this paper have been derived by using the {\sc Camb}~package \cite{camb}. JG acknowledges support from the Groupement d'Int\'er\^et Scientifique (GIS) "consortium Physique des 2 Infinis (P2I)". JM has been supported by the Polish Ministry of Science and Higher Education
Grant No. N203 386437.

\end{document}